\documentclass[aps,twocolumn,amssymb,amsmath,superscriptaddress]{revtex4-1}
\pdfoutput=1
\usepackage{bm}
\usepackage{color}
\usepackage{graphicx}
\usepackage{dcolumn}
\usepackage{graphicx}
\usepackage[colorlinks=true, letterpaper=true, pdfstartview=FitV, linkcolor=blue, citecolor=blue, urlcolor=blue]{hyperref}
\usepackage[normalem]{ulem}
\usepackage{amsmath}
\usepackage{epstopdf}
\usepackage{multirow}

\setcitestyle{square}

\makeatletter
\renewcommand\@biblabel[1]{#1.}
\makeatother

\begin{document}

\title{The first- and second-order magneto-optical effects and intrinsically anomalous transport in 2D van der Waals layered magnets Cr\textit{XY} (\textit{X} = S, Se, Te; \textit{Y} = Cl, Br, I)}

\author{Xiuxian Yang}
\affiliation{Centre for Quantum Physics, Key Laboratory of Advanced Optoelectronic Quantum Architecture and Measurement (MOE), School of Physics, Beijing Institute of Technology, Beijing, 100081, China}
\affiliation{Beijing Key Lab of Nanophotonics $\&$ Ultrafine Optoelectronic Systems, School of Physics, Beijing Institute of Technology, Beijing, 100081, China}
\affiliation{Kunming Institute of Physics, Kunming 650223, China}

\author{Ping Yang}
\affiliation{Centre for Quantum Physics, Key Laboratory of Advanced Optoelectronic Quantum Architecture and Measurement (MOE), School of Physics, Beijing Institute of Technology, Beijing, 100081, China}
\affiliation{Beijing Key Lab of Nanophotonics $\&$ Ultrafine Optoelectronic Systems, School of Physics, Beijing Institute of Technology, Beijing, 100081, China}

\author{Xiaodong Zhou}
\affiliation{Centre for Quantum Physics, Key Laboratory of Advanced Optoelectronic Quantum Architecture and Measurement (MOE), School of Physics, Beijing Institute of Technology, Beijing, 100081, China}
\affiliation{Beijing Key Lab of Nanophotonics $\&$ Ultrafine Optoelectronic Systems, School of Physics, Beijing Institute of Technology, Beijing, 100081, China}

\author{Wanxiang Feng}
\email{wxfeng@bit.edu.cn}
\affiliation{Centre for Quantum Physics, Key Laboratory of Advanced Optoelectronic Quantum Architecture and Measurement (MOE), School of Physics, Beijing Institute of Technology, Beijing, 100081, China}
\affiliation{Beijing Key Lab of Nanophotonics $\&$ Ultrafine Optoelectronic Systems, School of Physics, Beijing Institute of Technology, Beijing, 100081, China}

\author{Yugui Yao}
\affiliation{Centre for Quantum Physics, Key Laboratory of Advanced Optoelectronic Quantum Architecture and Measurement (MOE), School of Physics, Beijing Institute of Technology, Beijing, 100081, China}
\affiliation{Beijing Key Lab of Nanophotonics $\&$ Ultrafine Optoelectronic Systems, School of Physics, Beijing Institute of Technology, Beijing, 100081, China}

\date{\today}

\begin{abstract}
Recently, the two-dimensional magnetic semiconductor CrSBr has attracted considerable attention due to its excellent air-stable property and high magnetic critical temperature.  Here, we systematically investigate the electronic structure, magnetocrystalline anisotropy energy, first-order magneto-optical effects (Kerr and Faraday effects) and second-order magneto-optical effects (Sch{\"a}fer-Hubert and Voigt effects) as well as intrinsically anomalous transport properties (anomalous Hall, anomalous Nernst, and anomalous thermal Hall effects) of two-dimensional van der Waals layered magnets Cr\textit{XY} (\textit{X} = S, Se, Te; \textit{Y} = Cl, Br, I) by using the first-principles calculations.  Our results show that monolayer and bilayer Cr\textit{XY} (\textit{X} = S, Se) are narrow band gap semiconductors, whereas monolayer and bilayer CrTe\textit{Y} are multi-band metals.  The magnetic ground states of bilayer Cr\textit{XY} and the easy magnetization axis of monolayer and bilayer Cr\textit{XY} are confirmed by the magnetocrystalline anisotropy energy calculations.  Utilizing magnetic group theory analysis, the first-order magneto-optical effects as well as anomalous Hall, anomalous Nernst, and anomalous thermal Hall effects are identified to exist in ferromagnetic state with out-of-plane magnetization.  The second-order magneto-optical effects are not restricted by the above symmetry requirements, and therefore can arise in ferromagnetic and antiferromagnetic states with in-plane magnetization.  The calculated results are compared with the available theoretical and experimental data of other two-dimensional magnets and some conventional ferromagnets.  The present work reveals that monolayer and bilayer Cr\textit{XY} with superior magneto-optical responses and anomalous transport properties provide an excellent material platform for the promising applications of magneto-optical devices, spintronics, and spin caloritronics.
\end{abstract}

\maketitle
\setlength{\parskip}{1.8pt}

\section{Introduction}\label{intro}

According to the Mermin-Wagner theorem~\cite{Mermin1966}, the long-range magnetic order is strongly suppressed in low-dimensional systems mainly due to the enhanced thermal fluctuations, which makes the symmetry-breaking order unsustainable.  Nevertheless, recent experimental fabrications of atomically-thin ferromagnetic (FM) CrI$_3$~\cite{Huang2017} and Cr$_2$Ge$_2$Te$_6$~\cite{Gong2017} films, which are mechanically exfoliated from their bulk van der Waals (vdW) materials, indicated that the long-range magnetic order can be established in two-dimensional (2D) systems in which the magnetic anisotropy plays a vital role.  Since then, 2D magnetism has attracted enormous interest and more real materials were subsequently discovered, such as monolayer (ML) VSe$_2$~\cite{Bonilla2018,LIU2018419}, Fe$_3$GeTe$_2$~\cite{Fei2018,Deng2018}, and CrTe$_2$~\cite{ZhangXq2021}.  The ML VSe$_2$, Fe$_3$GeTe$_2$, and CrTe$_2$ are FM metals with relatively high Curie temperatures ($T_C$).  The $T_C$ of ML Fe$_3$GeTe$_2$ is around 130 K~\cite{Fei2018,Deng2018}, and the $T_C$ of ML VSe$_2$ and CrTe$_2$ are even close to room-temperature~\cite{Bonilla2018,LIU2018419,ZhangXq2021}, all of which are much larger than that of FM semiconductors CrI$_3$ ($T_C\sim$ 45 K)~\cite{Huang2017} and Cr$_2$Ge$_2$Te$_6$ ($T_C\sim$ 25 K)~\cite{Gong2017}.  From the aspect of practical applications for 2D spintronics, it is imperative to search for magnetic semiconductors with higher $T_C$.

A vdW layered material, CrSBr, was synthesized fifty-years ago~\cite{Hkatscher1966}, and its bulk form has been identified to be an A-type antiferromagnetic (AFM) semiconductor with the N{\'e}el temperature ($T_N$) of 132 K and electronic band gap of $\sim$1.5 eV~\cite{GOSER1990,Telford2020}.  Recently, several theoretical works predicted that ML CrSBr is a FM semiconductor with high $T_C$ ranging from 120 K to 300 K by different model calculations~\cite{Nicolas2018,Jiang2018,guoyl2018,WANG2019293,HanRl2020,HanRl2020,Wanghua2020,YangKe2021}.  Bulk CrSBr can be mechanically exfoliated to ML and  bilayer (BL), which were confirmed to be FM ($T_C$ $\sim$ 146 K) and AFM ($T_N$ $\sim$ 140 K), respectively, by a recent measurement of second harmonic generation~\cite{LeeKH2021}.  The ML and BL CrSBr have been paid much attention because their magnetic critical temperatures are obviously higher than that of CrI$_3$ and Cr$_2$Ge$_2$Te$_6$.  So far, some experimental works appear to focus on the fascinating properties of 2D CrSBr, including tunable magnetism~\cite{Telford2020,LeeKH2021,Nathan2021,telford2021hidden,rizzo2021visualizing,Cenker2022}, magnon-exciton coupling~\cite{bae2022}, interlayer electronic coupling~\cite{Nathan2021}, tunable electronic transport~\cite{telford2021hidden}, and etc.  The AFM semiconductor CrSBr was also used to introduce magnetism in graphene/CrSBr heterostructure by considering the proximity effect~\cite{Ghiasi2021}.  However, the magnetotransport properties of 2D CrSBr, such as magneto-optical Kerr~\cite{Kerr1877}, Faraday~\cite{Faraday1846}, Sch{\"a}fer-Hubert~\cite{SH1990}, and Voigt~\cite{voigt1908} effects (MOKE, MOFE, MOSHE, and MOVE) as well as anomalous Hall effect (AHE)~\cite{Nagaosa2010}, anomalous Nernst effect (ANE)~\cite{Nernst1887}, and anomalous thermal Hall effect (ATHE)~\cite{QinTao2011}, remain largely unexplored.  Since the family materials of Cr\textit{XY} (\textit{X} = S, Se, Te; \textit{Y} = Cl, Br, I) share the same crystal structure~\cite{WANG2019293}, atomically-thin films of Cr\textit{XY} should be easily obtained by the means of mechanical exfoliation, similarly to ML and BL CrSBr.  A systematic study on the magneto-optical effects (including MOKE, MOFE, MOSHE, and MOVE) and intrinsically anomalous transport properties (including AHE, ANE, and ATHE) of ML and BL Cr\textit{XY} will contribute to further understanding of this class of 2D vdW magnets.

In this work, using the first-principles calculations, we systemically investigate the electronic, magnetic, optical, magneto-optical, and anomalous transport properties of ML and BL Cr\textit{XY}.  We first reveal that CrS\textit{Y} and CrSe\textit{Y} are narrow band gap semiconductors, whereas CrTe\textit{Y} present multi-band metallic behaviors.  Then, the magnetic ground state and the easy axis of magnetization for Cr\textit{XY} are confirmed by magnetocrystalline anisotropy energy (MAE) calculations.  The results of MAE indicate that the easy axes of CrTeBr and CrTeI point along the out-of-plane direction (i.e., $z$-axis), while the other seven family members of Cr\textit{XY} prefer the in-plane magnetization.  The BL CrSI, CrSeCl, and CrTeCl show strong interlayer FM coupling, whereas other six family members exhibit interlayer AFM coupling.  Before calculating the magneto-optical and anomalous transport properties, we primarily analyze the shape of optical conductivity tensor by utilizing the magnetic group theory.  The uncovered symmetry requirements indicate that the first-order magneto-optical effects (MOKE and MOFE) as well as the AHE, ANE, and ATHE can only arise in the FM state with out-of-plane magnetization.  In order to explore the FM and AFM states with in-plane magnetization, we further study the second-order magneto-optical effects (MOSHE and MOVE).  The calculated first- and second-order magneto-optical effects as well as intrinsically anomalous transport properties of Cr\textit{XY} are compared with other 2D magnets and some conventional ferromagnets.  Our results show that 2D Cr\textit{XY} with superior magneto-optical responses and anomalous transport properties provide an excellent material platform for the promising applications of magneto-optical devices, spintronics, and spin caloritronics.

\begin{figure*}[htb]
	\includegraphics[width=2\columnwidth]{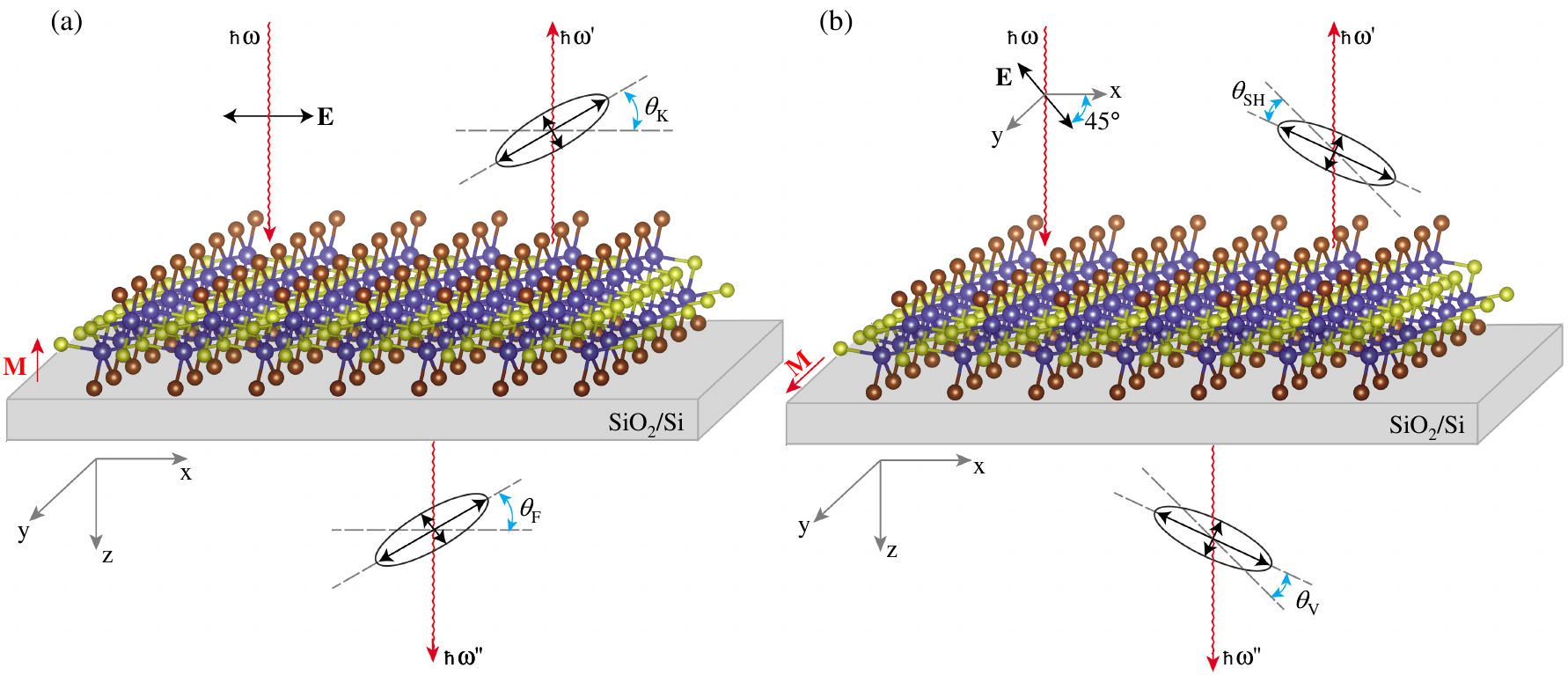}
	\caption{(Color online) Schematic illustration of magneto-optical Kerr and Faraday effects (a) as well as magneto-optical Sch{\"a}fer-Hubert and Voigt effects (b).  $\hbar\omega$, $\hbar\omega'$, and $\hbar\omega''$ indicate the incident, reflected, and transmission lights, respectively.  $\theta_\textnormal{K}$, $\theta_\textnormal{F}$, $\theta_\textnormal{SH}$, and $\theta_\textnormal{V}$ represent the Kerr, Faraday, Sch{\"a}fer-Hubert, and Voigt rotation angles, respectively.  \textbf{M} and \textbf{E} label the magnetization and electric field of incident light, respectively.  In (b), \textbf{E} and \textbf{M} form an angle of 45$^\circ$ on the $xy$ plane.}
	\label{fig:schematic}
\end{figure*}

\section{Theory and computational details}\label{method}

The magneto-optical effects, as a kind of fundamental magnetotransport phenomena in condensed matter physics, are usually considered to be a powerful probe of magnetism in low-dimensional systems~\cite{Huang2017,Gong2017}. The MOKE and MOFE can be described as the rotation of the polarization planes of reflected and transmitted lights after a linearly polarized light hits on the surface of magnetic materials, respectively.  Here, we only consider the MOKE and MOFE in the so-called polar geometry (e.g., both the directions of incident light and magnetization are along the $z$-axis), as shown in Fig.~\ref{fig:schematic}(a).  The MOKE and MOFE are commonly regarded as first-order magneto-optical effects as their magnitudes are linearly proportional to the strength of magnetization and their signs are odd in the direction of magnetization.  On the other hand, if the magnetization lies on the 2D atomic plane (i.e., $xy$-plane) and the angle between the electric field ($\textbf{E}$) of incident light and the direction of magnetization ($\textbf{M}$) is $45^\circ$, the polarization planes of reflected and transmitted lights still rotate, which are called MOSHE and MOVE [Fig.~\ref{fig:schematic}(b)], respectively.  The MOSHE and MOVE are considered to be second-order magneto-optical effects as their magnitudes are quadratic to the strength of magnetization and their signs are even in the direction of magnetization.  

It should be noted that in some FM and AFM materials, the first-order magneto-optical effects are prohibited due to the symmetry requirements, but the second-order magneto-optical effects can exist.  Here, we  take ML FM Fe$_3$GeTe$_2$ and BL AFM Fe$_3$GeTe$_2$ with in-plane magnetization~\cite{YangFGT2021} as examples to discuss this point.  The magnetic point group of ML FM Fe$_3$GeTe$_2$ with in-plane magnetization (when the spin points to the $x$-axis) is $m'm2'$, in which the mirror plane $\mathcal{M}_x$ is perpendicular to the spin and is parallel to the $z$-axis.  This mirror symmetry operation reverses the sign of the off-diagonal element of optical conductivity $\sigma_{xy}$ ($\equiv\sigma^{z}$), thus indicating $\sigma_{xy} = 0$.  If the spin points to the $y$-axis, the magnetic point group of ML FM Fe$_3$GeTe$_2$ is $m'm'2$, which contains a combined symmetry $\mathcal{TM}_z$ ($\mathcal{T}$ is the time-reversal symmetry, and $\mathcal{M}_z$ is a mirror plane perpendicular to the $z$-axis and parallel to the spin).  This combined symmetry operation reverses the sign of $\sigma_{xy}$, also suggesting $\sigma_{xy} = 0$.  For BL AFM Fe$_3$GeTe$_2$, the magnetic point groups are $2'/m$ and $2/m'$ when the spin points to the $x$- and $y$-axes, respectively.  Both the two magnetic point groups have a combined symmetry $\mathcal{TP}$ ($\mathcal{T}$ and $\mathcal{P}$ are the time-reversal and spatial inversion symmetries, respectively), which forces $\sigma_{xy}=0$.  Overall, in the cases of ML FM Fe$_3$GeTe$_2$ and BL AFM Fe$_3$GeTe$_2$ with in-plane magnetization, the vanishing off-diagonal element of optical conductivity ($\sigma_{xy}=0$) can not give rise to the first-order magneto-optical effects (MOKE and MOFE), refer to Eqs.~\eqref{eq:Kerr}--\eqref{eq:Faraday2}.  In contrast, the second-order magneto-optical effects (MOSHE and MOVE) depend on both the off-diagonal and diagonal elements, see Eqs.~\eqref{eq:SH}--\eqref{eq:Voigt}.  For the in-plane magnetization along either $x$- or $y$-axis, the two diagonal elements are not equivalent ($\sigma_{xx}\neq\sigma_{yy}$) due to the anisotropy of the in-plane optical response, which definitively induces the second-order magneto-optical effects regardless of whether the off-diagonal element is zero or not.

The Kerr and Faraday rotation angles ($\theta_\textnormal{K}$ and $\theta_\textnormal{F}$), which are the deflection of the polarization plane with respect to the incident light, can be used to quantitatively characterize the magneto-optical performance.  Additionally, the rotation angle $\theta_\textnormal{K(F)}$ and ellipticity $\varepsilon_\textnormal{K(F)}$ are usually combined into the complex Kerr (Faraday) angle~\cite{Suzuki1992,Guo1995,Ravindran1999},
\begin{equation}\label{eq:Kerr}
\phi_\textnormal{K}=\theta_\textnormal{K}+i\varepsilon_\textnormal{K}=i\frac{2\omega d}{c}\frac{\sigma_{xy}}{\sigma_{xx}^\textnormal{sub}},
\end{equation}
and 
\begin{equation}\label{eq:Faraday}
\phi_\textnormal{F}=\theta_\textnormal{F}+i\varepsilon_\textnormal{F}=\frac{\omega d}{2c}(n_+-n_-), 
\end{equation}
\begin{equation}\label{eq:Faraday2}
n_\pm^2=1+\frac{4\pi i}{\omega}(\sigma_{0}\pm i\sigma_{xy}),
\end{equation}
where $c$ is the speed of light in vacuum, $\omega$ is frequency of incident light, $d$ is the thickness of thin film, and $n_\pm$ are the refractive indices for the right- and left-circularly polarized light, respectively. $\sigma_{xx}^\textnormal{sub}$ is the diagonal element of the optical conductivity tensor for a nonmagnetic substrate (e.g., SiO$_2$/Si, as shown in Fig.~\ref{fig:schematic}).  In our calculations, only the interface between Cr\textit{XY} and SiO$_2$ layers is considered, and the SiO$_2$ layer is assumed to be thick enough.  In this case, the Si layer is unimportant as it does not influence the magneto-optical properties of Cr\textit{XY}.  The $\sigma_{xx}^\textnormal{sub}$ of the SiO$_2$ layer can be written as $i(1-n_\textnormal{sub}^2)\frac{\omega}{4\pi}$, here the $n_\textnormal{sub}$ is the energy-dependent refractive index of SiO$_2$ crystal (quartz)~\cite{GHOSH199995,n_sub}.  $\sigma_{xy}$ is the off-diagonal element of the optical conductivity tensor for a magnetic thin film and $\sigma_{0}=\frac{1}{2}(\sigma_{xx}+\sigma_{yy})$ due to in-plane anisotropy.

In the case of MOSHE and MOVE, the rotation angle $\theta_\textnormal{SH(V)}$ and ellipticity $\varepsilon_\textnormal{SH(V)}$ can be  also combined into the complex Sch{\"a}fer-Hubert and Voigt angles.  Considering the in-plane magnetization along the $y$-axis (Fig.~\ref{fig:schematic}), the complex Sch{\"a}fer-Hubert angle can be expressed as
\begin{eqnarray}\label{eq:SH}
\phi_\textnormal{SH}&=&\theta_\textnormal{SH}+i\varepsilon_\textnormal{SH}  \nonumber\\
&=& \frac{-i\omega d}{c(1-n_\textnormal{sub}^2)}(n_\parallel^2-n_\perp^2) 
\nonumber \\
&=& \frac{-i\omega d}{c(1-n_\textnormal{sub}^2)}[\epsilon_{yy}-\epsilon_{xx}-\frac{\epsilon_{zx}^2}{\epsilon_{zz}}],
\end{eqnarray}
and the complex Voigt angle is give by~\cite{Tesas2014,Mertins2001}
\begin{eqnarray}\label{eq:Voigt}
\phi_\textnormal{V}&=&\theta_\textnormal{V}-i\varepsilon_\textnormal{V}
\nonumber\\
&=&\frac{\omega d}{2ic}(n_\parallel-n_\perp)
\nonumber\\
&=&\frac{\omega d}{2ic}[\epsilon_{yy}^{\frac{1}{2}}-(\epsilon_{xx}+\frac{\epsilon_{zx}^2}{\epsilon_{zz}})^{\frac{1}{2}}].
\end{eqnarray}
Here, $\epsilon_{\alpha\beta}=\delta_{\alpha\beta}+\frac{4\pi i}{\omega}\sigma_{\alpha\beta}$ with ${\alpha,\beta}\in{x,y,z}$ is the permittivity tensor  (in cgs units), $n_\parallel$ and $n_\perp$ are the refractive indices of a magnetic thin film that are parallel and perpendicular to the direction of magnetization, respectively.

From Eqs.\eqref{eq:Kerr}--\eqref{eq:Voigt}, one would know that the key to calculate the first- and second-order magneto-optical effects is the optical conductivity, which can be obtained by the Kubo-Greenwood formula~\cite{Arash2008,Yates2007},
\begin{eqnarray}\label{eq:OPC}
\sigma_{xy}&=&\sigma_{xy}^1 (\omega) + i\sigma_{xy}^2 (\omega) \nonumber\\
&=& \frac{ie^2\hbar}{N_k V}\sum_{\textbf{k}}\sum_{n, m}\frac{f_{m\textbf{k}}-f_{n\textbf{k}}}{\varepsilon_{m\textbf{k}}-\varepsilon_{n\textbf{k}}} \nonumber\\
&&\times\frac{\langle\psi_{n\textbf{k}}|\hat{\upsilon}_x|\psi_{m\textbf{k}}\rangle\langle\psi_{m\textbf{k}}|\hat{\upsilon}_y|\psi_{n\textbf{k}}\rangle}{\varepsilon_{m\textbf{k}}-\varepsilon_{n\textbf{k}}-(\hbar\omega+i\eta)},
\end{eqnarray}
where the superscripts $1$ ($2$) indicates the real (imaginary) part of optical conductivity.  $\psi_{n\textbf{k}}$ and $\varepsilon_{n\textbf{k}}$ are the \textcolor{blue}{Bloch} function and interpolated energy at the band index $n$ and momentum $\textbf{k}$, respectively.  $f_{n\textbf{k}}$, \textcolor{blue}{$V$}, $N_k$, $\hat{v}_{x(y)}$, $\hbar\omega$, and $\eta$ are the Fermi-Dirac distribution function, volume of the unit cell, total number of $k$-points for sampling the Brillouin zone, the velocity operators, the photon energy, and energy smearing parameter, respectively.   $\eta$ was chosen to be 0.1 eV, corresponding to the carrier relaxation time of 6.5 fs, which is in the realistic range for multilayer transition metal chalcogenides~\cite{Strait2014,Sivadas2016}.  In this work, a sufficiently dense $k$-mesh of 450$\times$390$\times$1 is used to calculate the optical conductivity.  We also tested a denser $k$-mesh of 500$\times$430$\times$1 for the convergence of optical conductivity, see Figs.~\textcolor{blue}{S1(a)} and~\textcolor{blue}{S1(b)} in Supplemental Material~\cite{SuppMater}.

The AHE, which is featured by the transverse voltage drop induced by a longitudinal charge current in the absence of external magnetic field, is another fundamental magnetotransport phenomenon in condensed mater physics.  The contributions to AHE can be distinguished to extrinsic (skew-scattering and side-jump) and intrinsic mechanisms~\cite{Nagaosa2010}.  The extrinsic AHE is dependent on the scattering of electrons off impurities or due to disorder, while the intrinsic AHE originates from the Berry phase nature of the electrons in a perfect crystal.  In this work, we only focus on the intrinsically anomalous transport, i.e., intrinsic AHE.  The first-order magneto-optical effects and intrinsic AHE share the similar physical origin, and the dc limit of the real part of the off-diagonal element of optical conductivity, $\sigma_{xy}^1(\omega \rightarrow 0)$, is nothing but the intrinsic anomalous Hall conductivity (AHC).  Based on the linear response theory, the intrinsic AHC can be calculated~\cite{Yao2004},
\begin{equation}\label{eq:AHC}
\sigma_{xy}=-\frac{e^2\hbar}{V}\sum_{n,\textbf{k}}f_{n\textbf{k}}\Omega^n_{xy}(\textbf{k}),
\end{equation}
in which $f_{n\textbf{k}}$ is the Fermi-Dirac distribution function, $V$ is volume of the unit cell, and $\Omega^n_{xy}(\textbf{k})$ is the band- and momentum-resolved Berry curvatures, given by
\begin{equation}\label{eq:Berry}
\Omega^n_{xy}(\textbf{k})=-\sum_{n'\neq n}\frac{2 {\rm Im}[\langle\psi_{n\textbf{k}}|\hat{v}_x|\psi_{n'\textbf{k}}\rangle\langle\psi_{n'\textbf{k}}|\hat{v}_y|\psi_{n\textbf{k}}\rangle]}{(\varepsilon_{n\textbf{k}}-\varepsilon_{n'\textbf{k}})^2}.
\end{equation}
To calculate $\sigma_{xy}$, a $k$-mesh of 600$\times$520$\times$1 was used.  We also tested the convergence of $\sigma_{xy}$ on a denser $k$-mesh of 750$\times$650$\times$1, see supplemental Fig.~\textcolor{blue}{S1(c)}~\cite{SuppMater}.

In addition to the AHE, the transverse charge current can also be induced by a longitudinal temperature gradient field, named ANE~\cite{Nernst1887}.  Additionally, a transverse thermal current arise under the longitudinal temperature gradient field, which is called ATHE or anomalous Righi-Leduc effect~\cite{QinTao2011}.  The magnetic materials that show large anomalous Nernst conductivity (ANC) and anomalous thermal Hall conductivity (ATHC) have great potentials to apply for spin caloritronics or thermoelectric devices.  The AHE, ANE, and ATHE are closely related to each other, and can be expressed through the anomalous transport coefficients in the generalized Landauer-B\"uttiker formalism~\cite{ashcroft1976solid,Houten1992,behnia2015fundamentals},
\begin{equation}\label{eq:LB}
R^{(n)}_{xy}=\int^\infty_{-\infty}dE(E-\mu)^n(-\frac{\partial f}{\partial E})\sigma_{xy}(E),
\end{equation}
where $E$, $\mu$, and $f$ are energy, chemical potential, and Fermi-Dirac distribution function, respectively.  $\sigma_{xy}$ is the intrinsic AHC calculated by Eq.~\eqref{eq:AHC} at zero temperature.  Then, the temperature-dependent ANC ($\alpha_{xy}$) and ATHC ($\kappa_{xy}$) are written as
\begin{equation}\label{eq:ANC}
\alpha_{xy}=-R^{(1)}_{xy}/eT,
\end{equation}
\begin{equation}\label{eq:ATHC}
\kappa_{xy}=R^{(2)}_{xy}/e^2T.
\end{equation}
To calculate $\alpha_{xy}$ and $\kappa_{xy}$, the integral in Eq.~\eqref{eq:LB} is set in the range of [-1.0, 2.5] eV for semiconductors and [-1.0, 1.0] eV for metals, with the energy interval of 0.0001 eV.

The first-principles density functional theory calculations were performed by utilizing the Vienna \textit{ab initio} simulation package (\textsc{vasp})~\cite{Kresse1993,Kresse1996}.  The projector augmented wave (PAW) method~\cite{Blochl1994} was used and the generalized gradient approximation (GGA) with the Perdew-Burke-Ernzerhof parameterization~\cite{Perdew1996} was adopted to treat the exchange-correlation functional.  Dispersion correction was performed at the van der Waals density functional (vdW-DF) level~\cite{Klime2009,LeeK2010,Klime2011}, where the optB86b functional was used for the exchange potential.  For the ML and BL structures, the vacuum layer with a thickness of at least 15 {\AA} was used to avoid the interactions between adjacent layers.  All structures were fully relaxed until the force on each atom was less than 0.001 eV$\cdot${\AA$^{-1}$}.  To describe the strong correlation effect of the $d$ orbitals on Cr atom, the GGA+U method~\cite{Liecht1995} was used by taking appropriate parameters of U and J in the range of 4.05 $\sim$ 4.40 eV and 0.80 $\sim$ 0.97 eV~\cite{WANG2019293}, respectively.  The values of U and J  used for each compound are listed in supplemental Tab.~\textcolor{blue}{S1}~\cite{SuppMater}.  To obtain accurate MAE, a large plane-wave cutoff energy of 600 eV and a Monkhorst-Pack $k$-mesh of 15$\times$13$\times$1 were used.  The spin-orbit coupling (SOC) was included in the calculations of MAE, magneto-optical effects, and anomalous transport properties.  The maximally localized Wannier functions were constructed by \textsc{wannier90} package~\cite{Arash2008}, and the $d$, $p$, and $p$ orbitals of Cr, \textit{X} (\textit{X} = S, Se, Te), and \textit{Y} (\textit{Y} = Cl, Br, I) atoms were projected onto the Wannier functions, respectively.  Totally, there are 44 and 88 Wannier functions for ML and BL Cr\textit{XY}, respectively, which converged adequately the results of magneto-optical effects and anomalous transport properties.  Similarly, the computational technique employing the maximally localized  Wannier functions can be in principle used to calculate the physical quantities that are closely related to orbital angular momenta~\cite{Wozniak2020,Thorsten2020,Jonathan2020,Xuan2020}.  The \textsc{isotropy} software~\cite{isotropy} was used to identify magnetic space and point groups.  The workflow of the various codes used in our work is depicted in supplemental Fig.~\textcolor{blue}{S2}~\cite{SuppMater}.

\section{Results and discussion}\label{results}

\subsection{Crystal, magnetic, and electronic structures}
\begin{figure}
	\includegraphics[width=1\columnwidth]{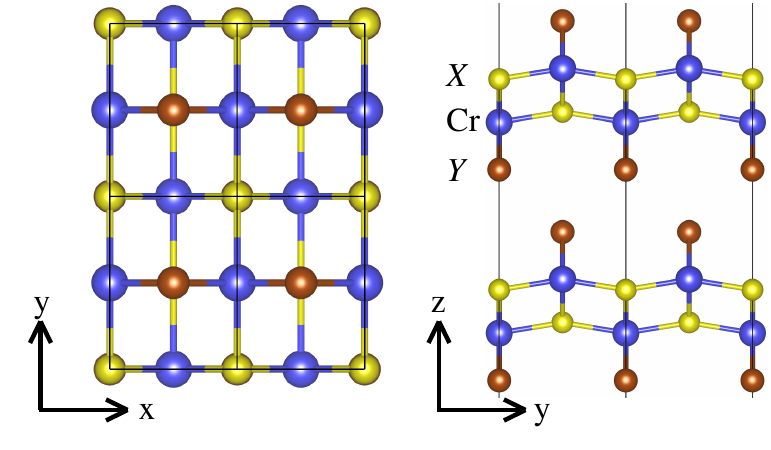}
	\caption{(Color online) Top and side views of layered Cr\textit{XY} (\textit{X} = S, Se, Te; \textit{Y} = Cl, Br, I).  The blue, yellow, and brown spheres represent Cr, \textit{X}, and \textit{Y} atoms, respectively.  The black solid lines draw out the primitive cell.  The experimental stacking order of bilayer structure is adopted.}
	\label{fig:crystall}
\end{figure}

Figure~\ref{fig:crystall} shows the crystal structure of vdW layered materials Cr\textit{XY} (\textit{X} = S, Se, Te; \textit{Y} = Cl, Br, I).  One can see that Cr\textit{XY} has a rectangular 2D primitive cell, in which a ML is made of two buckled planes of Cr$X$ sandwiched by two $Y$ sheets, and a BL is formed by two MLs with the AA stacking order referring to experimental works~\cite{Telford2020,Nathan2021,LeeKH2021,Klein2021}.  The relaxed lattice constants of ML and BL structures are collected in supplemental Tab.~\textcolor{blue}{S1}~\cite{SuppMater}.  The calculated lattice constants of ML CrSBr, $a$ = 3.54 {\AA}, $b$ = 4.79 {\AA}, and thickness $d$ = 8.00 {\AA}, are in good agreement with the experimental results, $a$ = 3.50 {\AA}, $b$ = 4.76 {\AA}~\cite{Telford2020}, and $d$ = 7.8 $\pm$ 0.3 {\AA}~\cite{LeeKH2021}.

The magnetic anisotropy plays a vital role in stabilizing the long-range magnetic order of 2D systems.  The MAE is defined as the difference of total energies between different magnetization directions (such as, along the $x$-, $y$-, and $z$-axis), which can be directly obtained by the first-principles calculations.  To determine the magnetic ground states, we calculate the MAE of ML and BL Cr\textit{XY}.  Since the FM ground state of ML Cr\textit{XY} was predicted by a recent theoretical work~\cite{WANG2019293} and has been confirmed by two experimental works~\cite{Nathan2021,LeeKH2021}, here we directly adopt the FM configuration as the ground state of ML Cr\textit{XY}.  Thus, the magnetic ground states for ML Cr\textit{XY} can also be determined by the Goodenough-Kanamori rules, i.e., superexchange mechanism~\cite{Anderson1950,GOODENOUGH1958,KANAMORI195987}.  The magnetic exchange interactions can be judged by the angle of the chemical bonds connecting the ligand and two nearest-neighboring magnetic atoms. Particularly, a system prefers to be ferromagnetic if the angle equals 90$^\circ$.  Taking ML CrSBr as an example, the bond angle of Cr-Br-Cr is 88.96$^\circ$, and the bond angles of Cr-S-Cr are 94.24$^\circ$ and 96.58$^\circ$, suggesting the ferromagnetic ground state.  Our calculated MAE indicate that ML CrSI, CrSe\textit{Y} prefer to the in-plane magnetization along the $x$-axis, ML CrSCl, CrSBr, and CrTeCl are in favor of in-plane magnetization along the $y$-axis, while ML CrTeBr and CrTeI take for the out-of-plane $z$-axis magnetization.  The evolution of the easy axis from in-plane to out-of-plane direction can be explained by the SOC strength of the systems.  The atomic SOC strength increases gradually in the consequences of S $\rightarrow$ Se $\rightarrow$ Te and Cl $\rightarrow$ Br $\rightarrow$ I, and CrTeBr and CrTeI have the almost largest SOC strength among the other family compounds.  It leads to the MAE results that CrTeBr and CrTeI have the out-of-plane $z$-axis magnetization, while the other family compounds have the in-plane $x$- or $y$-axis magnetization.  This is similar to the case of 2D magnetic materials Cr$X_3$ ($X$ = Cl, Br, I)~\cite{KimHH2019}, in which the magnetization direction changes from in-plane (CrCl$_3$) to out-of-plane (CrI$_3$) with the increasing of SOC strength.  For the BL Cr\textit{XY}, we compare the total energy of FM and AFM states (see supplemental Fig.~\textcolor{blue}{S3}~\cite{SuppMater}), and find that CrSI, CrSeCl, and CrTeCl exhibit the FM ground state, while the other six family members prefer to be AFM state.  The easy magnetization axes of BL Cr\textit{XY} are identical to that of ML Cr\textit{XY}.  The above results are summarized in supplemental Tab.~\textcolor{blue}{S1}~\cite{SuppMater}.  The easy magnetization axis of CrSBr is along the $y$-axis, which is in an accordance with the experimental reports~\cite{Telford2020,Nathan2021,LeeKH2021}.

\begin{figure}
	\includegraphics[width=1\columnwidth]{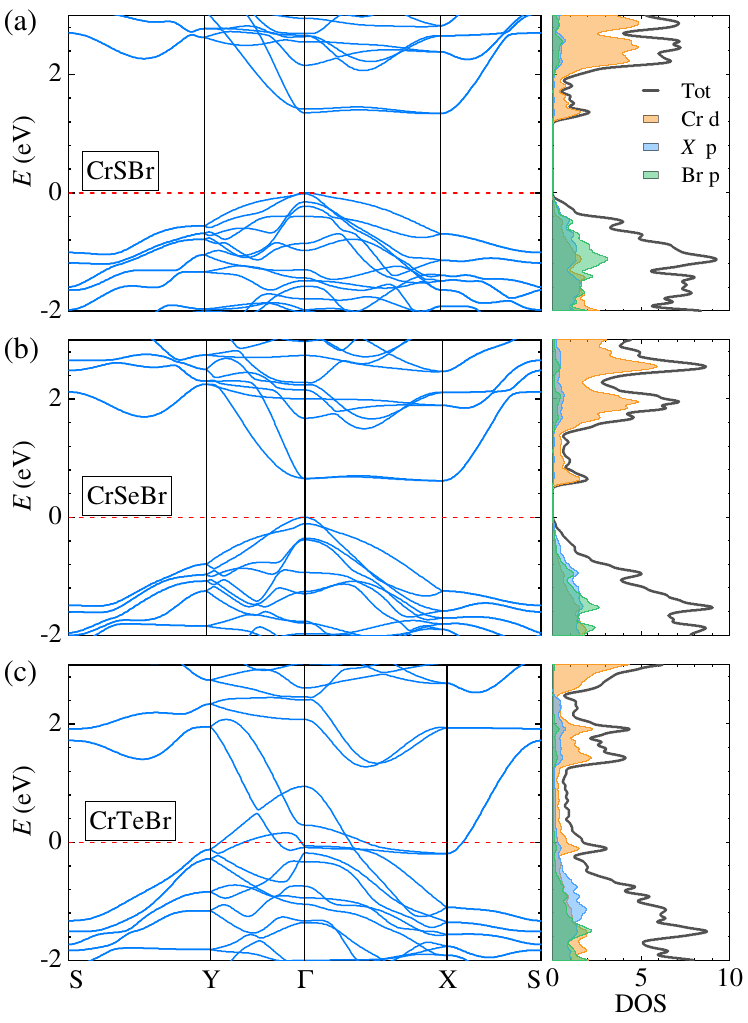}
	\caption{(Color online)  The relativistic band structures and orbital-decomposed density of states (in unit of states/eV per cell) for monolayer ferromagnetic Cr\textit{X}Br (\textit{X} = S, Se, Te) with out-of-plane magnetization.}
	\label{fig:band}
\end{figure}

After obtaining the magnetic ground states of Cr\textit{XY}, we shall further ascertain the corresponding magnetic groups as the group theory is a powerful tool to figure out the symmetry requirements for the magneto-optical effects and anomalous transport properties.  The off-diagonal elements of optical conductivity tensor [$\sigma_{\alpha\beta}(\omega)$] fully determine the symmetry requirements of first-order magneto-optical effects [see Eqs.~\eqref{eq:Kerr}--\eqref{eq:Faraday2}].  At zero frequency limit, $\sigma_{\alpha\beta}(\omega\rightarrow0)$ is nothing but the intrinsic AHC [i.e., Eq.~\eqref{eq:AHC}], and therefore the symmetry requirements of AHE, ANE, and ATHE [refer to Eqs.~\eqref{eq:LB}--\eqref{eq:ATHC}] should be the same as that of $\sigma_{\alpha\beta}(\omega)$.  Here, we only need to analyze the symmetry results of $\sigma_{\alpha\beta}(\omega)$ under relevant magnetic groups.

Since the off-diagonal elements of optical conductivity tensor can be regarded as a pseudovector (like spin), its vector-form notation, $\boldsymbol{\sigma} = [\sigma^x, \sigma^y, \sigma^z] = [\sigma_{yz}, \sigma_{zx}, \sigma_{xy}]$, is used here for convenience.  In 2D systems, only $\sigma^z$ ($\equiv\sigma_{xy}$) is potentially nonzero, while $\sigma^x$ ($\equiv\sigma_{yz}$) and $\sigma^y$ ($\equiv\sigma_{zx}$) are always zero.  It can be easily understood from Eq.~\eqref{eq:OPC} as the zero velocity along the $z$-axis (i.e., $\hat{v}_z=0$, meaning that the electrons can not move along the out-of-plane direction in 2D systems) definitely gives rise to vanishing $\sigma_{yz}$ and $\sigma_{zx}$.  For ML and BL FM Cr\textit{XY} with any magnetization directions ($x$-, $y$-, or $z$-axis), the magnetic space and point groups of are always identified to be $Pm'm'n$ and $m'm'm$ [$D_{2h}(C_{2h})$ in Sch{\"o}nflies notation], respectively.  It is sufficient for us to analyze magnetic point group due to the translationally invariance of $\sigma_{xy}$~\cite{XD-Zhou2019}.  Hence, we only discuss the magnetic point group of $m'm'm$, in which there are one mirror plane $\mathcal{M}$ and two combined symmetries $\mathcal{TM}$ ($\mathcal{T}$ is the time-reversal symmetry).  For the in-plane magnetization along the $x$-axis ($y$-axis), the mirror plane is $\mathcal{M}_x$ ($\mathcal{M}_y$), which is perpendicular to the spin magnetic moment, such a mirror operation reverses the sign of $\sigma_{xy}$, indicating that $\sigma_{xy}$ should be zero.  On the other hand, for out-of-plane magnetization, the mirror plane is $\mathcal{M}_z$ and two combined symmetries are $\mathcal{TM}_x$ and $\mathcal{TM}_y$.  The spin magnetic moment is perpendicular to the $\mathcal{M}_z$, which does not change the sign of $\sigma_{xy}$.  Moreover, the mirror planes $\mathcal{M}_x$ and $\mathcal{M}_y$ are parallel to the spin, and both $\mathcal{TM}_x$ and $\mathcal{TM}_y$ preserve the sign of $\sigma_{xy}$.  A conclusion here is that $\sigma_{xy}$ is nonzero (zero) for ML and BL FM Cr\textit{XY} with out-of-plane (in-plane) magnetization.  In the case of the BL AFM Cr\textit{XY}, the magnetic space and point groups are $Pm'm'n'$ and $m'm'm'$ [$D_{2h}(D_{2})$ in Sch{\"o}nflies notation] for out-of-plane magnetization, and are $Pm'mn$ and $mmm'$ [$D_{2h}(C_{2v})$ in Sch{\"o}nflies notation] for in-plane magnetization.  There is a combined symmetry $\mathcal{TP}$ ($\mathcal{P}$ is the spatial inversion) in the two point groups of $m'm'm'$ and $mmm'$, which forces $\sigma_{xy}$ to be zero.  Overall, a nonzero $\sigma_{xy}$ can only exist in ML and BL FM Cr\textit{XY} with out-of-plane magnetization, which turn out to be our target systems for studying the MOKE, MOFE as well as AHE, ANE, and ATHE.  It should be noted that the out-of-plane magnetization is not the magnetic ground states for some family members of Cr\textit{XY}, however, an applied external magnetic filed can easily tune the spin direction, as already realized in layered CrSBr~\cite{Nathan2021}.

\begin{figure*}
	\includegraphics[width=1.75\columnwidth]{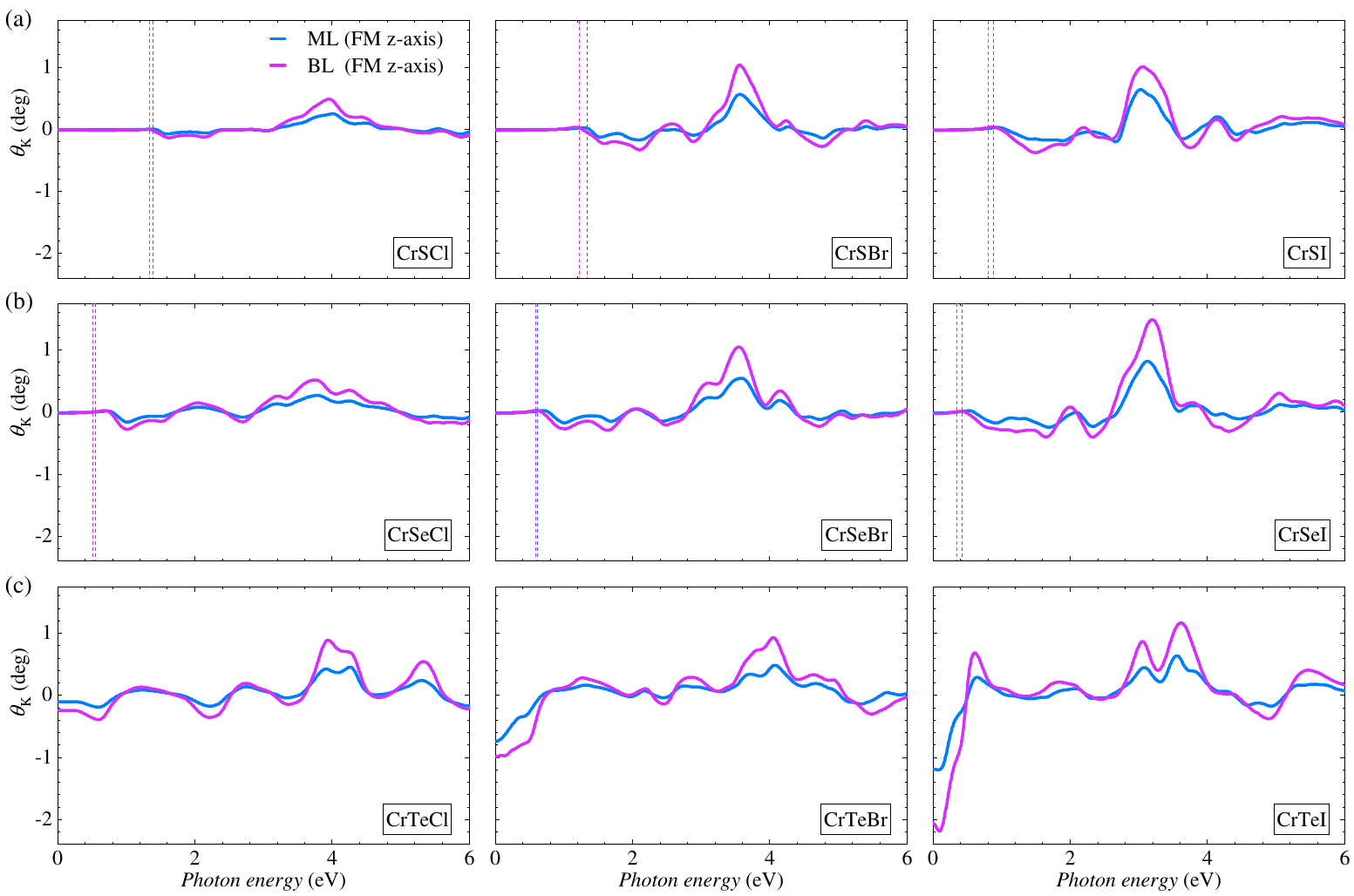}
	\caption{(Color online) The magneto-optical Kerr rotation angles ($\theta_K$) of monolayer (blue solid lines) and bilayer (pink solid lines) ferromagnetic Cr\textit{XY} (\textit{X} = S, Se, Te; \textit{Y} = Cl, Br, I) with out-of-plane magnetization.  The vertical dashed lines in (a) and (b) indicate the band gaps.}
	\label{fig:MOK}
\end{figure*}

\begin{figure*}[htbp]
	\includegraphics[width=1.75\columnwidth]{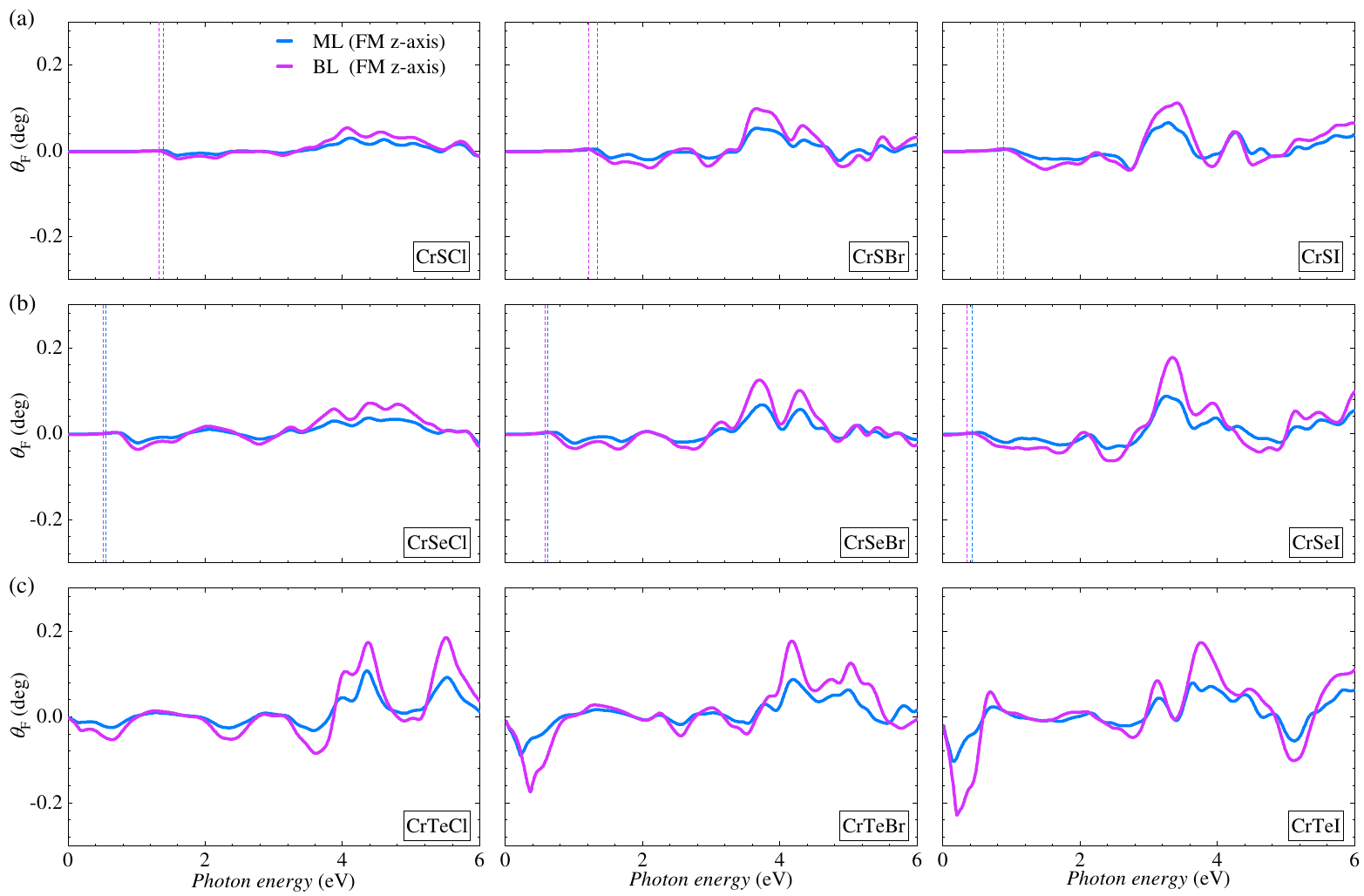}
	\caption{(Color online)  The magneto-optical Faraday rotation angles ($\theta_F$) of monolayer (blue solid lines) and bilayer (pink solid lines) ferromagnetic Cr\textit{XY} (\textit{X} = S, Se, Te; \textit{Y} = Cl, Br, I) with out-of-plane magnetization. The vertical dashed lines in (a) and (b) indicate the band gaps.}
	\label{fig:MOF}
\end{figure*}

\begin{figure}
	\includegraphics[width=0.9\columnwidth]{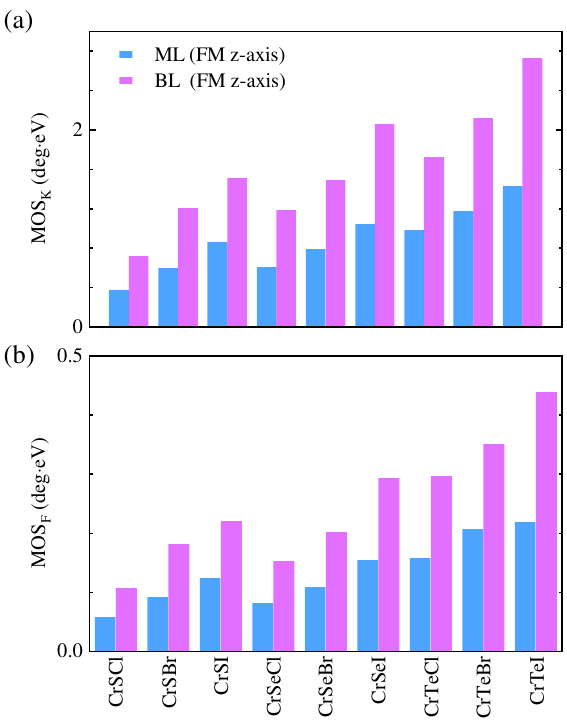}
	\caption{(Color online)  The magneto-optical strength of Kerr (a) and Faraday (b) effects for monolayer and bilayer Cr\textit{XY} with out-of-plane magnetization.}
	\label{fig:MOS}
\end{figure}

The first-order magneto-optical effects (MOKE and MOFE) can not exist in FM and AFM Cr\textit{XY} with in-plane magnetization, but it is not the case for the second-order magneto-optical effects (MOSHE and MOVE).  According to the Onsager relations, the diagonal elements of the permittivity tensor are even in magnetization (\textbf{M}), e.g., $\epsilon_{yy}(-\textbf{M})=\epsilon_{yy}(\textbf{M})$.  Furthermore, equations~\eqref{eq:SH} and~\eqref{eq:Voigt} show that the complex Sch{\"a}fer-Hubert and Voigt angles depend on square of $\epsilon_{zx}$.  Therefore, the MOSHE and MOVE must be even in \textbf{M}.  In the case of FM and AFM Cr\textit{XY} with in-plane magnetization, although $\epsilon_{zx}$ is always zero, the MOSHE and MOVE do exist because of the nonvanishing $\epsilon_{yy}$ and $\epsilon_{xx}$ (the relation of $\epsilon_{yy}\neq\epsilon_{xx}$ also satisfies).  Owing to this, we use the MOSHE and MOVE to characterize the in-plane magnetized ML FM and BL FM/AFM Cr\textit{XY}.  The complex Sch{\"a}fer-Hubert and Voigt angles are calculated by orienting the spins along the $y$-axis, and the results of $x$-axis in-plane magnetization can be simply obtained by  putting a minus sign, refer to Eqs.~\eqref{eq:SH} and~\eqref{eq:Voigt}.

The relativistic band structures of ML FM and BL FM/AFM Cr\textit{XY} with out-of-plane and in-plane magnetization are plotted in supplemental Figs.~\textcolor{blue}{S4}--\textcolor{blue}{S9}~\cite{SuppMater}.  The results show that CrS\textit{Y} and CrSe\textit{Y} are narrow band gap semiconductors, while CrTe\textit{Y} are multi-band metals.  All the band gaps of CrS\textit{Y} and CrSe\textit{Y} are summarized in supplemental Tab.~\textcolor{blue}{S1}~\cite{SuppMater}.  Here, we choose ML FM Cr\textit{X}Br (\textit{X} = S, Se, Te) as representative examples to discuss the electronic properties of Cr\textit{XY} family materials, as shown in Fig.~\ref{fig:band}.  A clear trend can be seen that the band gaps of Cr\textit{X}Br become smaller and eventually disappear with the increasing of atomic number (e.g., S $\rightarrow$ Se $\rightarrow$ Te).  The orbital-decomposed density of states of ML FM Cr\textit{X}Br show that the $d$ orbitals of Cr atom and $p$ orbitals of \textit{X} and Br atoms are dominant components near the Fermi level.  For CrSBr, the calculated band gaps of ML FM,  BL FM, and BL AFM states are 1.34 eV, 1.22 eV, and 1.30 eV, respectively, which are in good agreement with that of bulk material ($\sim$ 1.25 eV)~\cite{Telford2020}.  Since CrS\textit{Y} and CrSe\textit{Y} are narrow band gap semiconductors, the electrons and holes could bound in pairs to form quasiparticle excitons.  Taking CrSBr as an example~\cite{Nathan2021}, the exciton binding energies for ML FM, BL FM, and BL AFM states are 0.5 eV, 0.37 eV, and 0.46 eV, respectively, which are apparently smaller than the exciton binding energies of ML FM CrI$_3$ (1.7 eV~\cite{Wumeng2020}, 1.06 eV~\cite{Alejandro2020}, and 0.84~\cite{Alejandro2020}).   It indicates that the relatively weak excitonic effects in CrS\textit{Y} and CrSe\textit{Y} may not significantly influence the optical and magneto-optical properties, in contrast to CrI$_3$~\cite{Wumeng2020,Alejandro2020}.

\begin{table*}[htpb]\footnotesize
	\caption{The maximums of Kerr, Faraday, Sch$\ddot{a}$fer-Hubert, and Voigt rotation angles ($\theta_K$, $\theta_F$, $\theta_{SH}$, and $\theta_{V}$) in the moderate energy range (1 $\sim$ 5 eV) as well as the corresponding incident photon energy ($\hbar\omega$) for monolayer and bilayer Cr\textit{XY}.  Here, $\theta_K$ and $\theta_F$ are calculated under the ferromagnetic states with out-of-plane magnetization, while $\theta_{SH}$ and $\theta_{V}$ are calculated under ferromagnetic or antiferromagnetic states with in-plane $y$-axis magnetization.  For a better comparison, $\theta_F$ and $\theta_{V}$ are given in two units, deg and deg/$\mu$m (in the parentheses), which can be converted to each other by considering the effective thickness of crystal structures along the $z$-axis.}
	\label{tab:MOKF}
	\begin{ruledtabular}
	\begingroup
	\setlength{\tabcolsep}{4.5pt} 
	\renewcommand{\arraystretch}{1.5} 
	\begin{tabular}{llccllclcl}
		
		& &$\theta_K$ (deg) &$\hbar\omega$ (eV) &$\theta_F$ (deg) (deg/$\mu$m) &$\hbar\omega$ (eV) &$\theta_{SH}$ (deg) &$\hbar\omega$ (eV) &$\theta_V$ (deg) (deg/$\mu$m) &$\hbar\omega$ (eV)\\
		\hline
		CrSCl   &ML (FM)  &0.25 &4.01  &0.04 (54.29)&4.11  &-4.65&3.65   &0.41 (559.34)&3.69 \\          
		&BL (FM) &0.48 &3.95  &0.06 (33.93)&4.07  &-9.30&3.66 &0.85 (579.39) &3.70\\	 
		&BL (AFM)  &  &  &  &  &-8.64 &3.66   &0.76 (513.05) &3.70\\
		\hline		
		CrSBr   &ML (FM)  &0.56&3.56  &0.06 (75.02)&3.66  &-3.92&3.10   &0.40 (496.28)&3.13\\          
		&BL (FM) &1.04&3.56  &0.11 (68.76)&3.65  & 6.01&3.69   &-0.80 (-502.03) &2.66\\
		&BL (AFM)  &  &  &  &  &9.87&2.24  &-0.77 (-478.74) &2.28\\
		\hline			
		CrSI    &ML (FM) &0.64&3.02  &0.08 (90.60)&3.28  &5.90&3.30  &-0.55 (-624.74) &3.37 \\          
		&BL (FM) &1.00&3.06  &0.12 (67.95)&3.41  &8.63&3.37  &-0.90 (-507.99) &3.41\\
		&BL (AFM)  &  &  &  &  &6.60&3.35  &-0.77 (-436.75) &2.30\\
		\hline	
		CrSeCl  &ML (FM)  &0.27&3.78  &0.04 (49.52)&4.38  &-3.66&3.40  &0.31 (421.07)&4.76\\          
		&BL (FM) &0.51&3.74  &0.08 (54.81)&4.41  &6.99&2.82   &-0.84 (-577.27) &2.91\\
		&BL (AFM)  &  &  &  &  &7.00&2.74  &-0.76 (-524.16) &2.89\\
		\hline	
		CrSeBr  &ML (FM)  &0.54&3.58  &0.08 (101.94)&3.74  &3.93&2.30  &0.43 (552.31) &1.65\\          
		&BL (FM) &1.05&3.55  &0.15 (95.57)&3.70  &7.48&2.29  &0.76 (484.44) &1.63\\
		&BL (AFM)  &  &  &  &  &10.27&2.45   &0.87 (551.30) &1.63\\
		\hline	
		CrSeI   &ML (FM)  &0.82&3.12  &0.10 (114.72)&3.25  &3.62&1.90  &-0.44 (-504.61)&1.93\\          
		&BL (FM) &1.49&3.19  &0.20 (114.72)&3.35  &6.27&2.00   &0.63 (362.78) &1.27\\
		&BL (AFM)  &  &  &  &  &5.17 &1.92   &-0.53 (-304.59) &1.99\\ 
		\hline	
		CrTeCl  &ML (FM)  &0.45&4.26  &0.10 (132.58)&4.36  &8.54&2.66  &-1.01 (-1342.27)&2.72\\          
		&BL (FM)   &0.89&3.94  &0.17 (114.92)&4.37  &13.39&2.76  &-1.67 (-1107.05) &2.85\\
		&BL (AFM)  &  &  &  &  &11.31 &2.77  &-1.50 (-994.53) &2.91\\
		\hline	
		CrTeBr  &ML (FM)  &0.48&4.07  &0.09 (110.91)&4.19  &4.68&2.45   &-0.57 (-731.54)&2.50\\          
		&BL (FM)   &0.93&4.05  &0.18 (112.56)&4.17  &8.08&2.54  &-1.07 (-680.64) &2.61\\
		&BL (AFM)  &  &  &  &  &10.15 &2.66  &-1.27 (-811.67) &2.70\\
		\hline	
		CrTeI   &ML (FM)  &0.64&3.56  &0.08 (93.56)&3.64  &-2.53&4.28  &0.32 (377.20)&4.48\\          
		&BL (FM)   &1.17&3.61  &0.19 (112.48)&3.76  &-4.94&3.40  &-0.46 (-269.45) &2.48\\
		&BL (AFM)  &  &  &  &  &-7.25 &3.39  &-0.66 (-388.03) &3.05\\
		
	\end{tabular}
	\endgroup
\end{ruledtabular}
\end{table*}

\begin{table*}[htpb]\footnotesize
	\caption{The magneto-optical Kerr, Faraday, Sch$\ddot{a}$fer-Hubert, and Voigt rotation angles ($\theta_K$, $\theta_F$, $\theta_{SH}$, and $\theta_{V}$) for Cr\textit{XY} and other 2D magnetic materials.  For a better comparison, the absolute values of rotation angles are used here as the sign only indicates the rotation direction with respect to the principle axis of incident light that may have different definitions in literatures.  Moreover, the rotation angles are given in a range due to the dependence on the number of layers for some 2D materials.  The data marked by "Exp" are experimental measurements, while the others are computational results.}
	\label{tab:Comparison}
	\begin{ruledtabular}
		\begingroup
		\setlength{\tabcolsep}{4.5pt} 
		\renewcommand{\arraystretch}{1.5} 
		\begin{tabular}{llccc}
			
			&$\theta_K$ (deg) &$\theta_F$(deg/$\mu$m) &$\theta_{SH}$ (deg)  &$\theta_V$  (deg) \\
			\hline
			Cr\textit{XY}    &0.25$\sim$1.49  &33.93$\sim$132.58  &2.53$\sim$13.39 &0.31$\sim$1.67\\          
			\hline
			CrI$_3$ &0.29$\sim$2.86~\cite{Huang2017}$^{\textnormal{Exp}}$  & 50$\sim$108~\cite{Kumar2019}\\
			&0.60$\sim$1.30~\cite{Kumar2019}  \\
			\hline 
			CrTe$_2$ & 0.24$\sim$1.76~\cite{YangXX2021} &30$\sim$173~\cite{YangXX2021}\\
			\hline
			CrGeTe$_3$ &0.0007$\sim$0.002~\cite{Gong2017}$^{\textnormal{Exp}}$ &100$\sim$120~\cite{FangY2018}\\
			&0.9$\sim$2.2~\cite{FangY2018} \\
			\hline
			GA & 0.13$\sim$0.81~\cite{Zhou2017} &9.3$\sim$137.8~\cite{Zhou2017}\\
			\hline
			BP &0.02$\sim$0.12~\cite{Zhou2017} &1.4$\sim$18.9~\cite{Zhou2017}\\
			\hline
			InS &0.34~\cite{Feng2017} &84.3~\cite{Feng2017}\\
			\hline
			Fe$_n$GeTe$_2$ ($n$ = 3, 4, 5) &0.74$\sim$2.07~\cite{YangFGT2021} &50.28 $\sim$ 222.66~\cite{YangFGT2021} \\
			\hline
			CoO & & &0.007~\cite{ZhengZ2018}$^{\textnormal{Exp}}$ \\
			\hline
			CuMnAs & & & &0.023~\cite{Saidl2017}$^{\textnormal{Exp}}$

		\end{tabular}
		\endgroup
	\end{ruledtabular}
\end{table*}

\subsection{First-order magnetic-optical effects}

The calculated optical conductivities ($\sigma_{xx}$, $\sigma_{yy}$, and $\sigma_{xy}$ ) for ML and BL FM Cr\textit{XY} with out-of-plane magnetization are displayed in supplemental Figs.~\textcolor{blue}{S10}--\textcolor{blue}{S13}~\cite{SuppMater}.  The spectra of optical conductivity of ML and BL structures have similar shape due to the weak vdW interaction~\cite{FangY2018}.  After obtaining the optical conductivity tensors, we can calculate the first-order magneto-optic effects, i.e., MOKE and MOFE, according to Eqs.~\eqref{eq:Kerr}--\eqref{eq:Faraday2}.  The Kerr and Faraday rotation angles ($\theta_K$ and $\theta_F$) and the corresponding ellipticities ($\varepsilon_{K}$ and $\varepsilon_{F}$) as a function of photon energy are illustrated in Figs.~\ref{fig:MOK}--\ref{fig:MOF} and supplemental Figs.~\textcolor{blue}{S14}--\textcolor{blue}{S15}~\cite{SuppMater}, respectively.  From Figs.~\ref{fig:MOK}(a)--\ref{fig:MOK}(b), one can observe that the $\theta_K$ of ML and BL CrS\textit{Y} and CrSe\textit{Y} are vanishing within the band gap due to the semiconducting nature, while increase largely in high energy region.  The ML and BL CrTe\textit{Y} exhibit distinct $\theta_K$ from zero frequency, suggesting the metallic property, as displayed in Fig.~\ref{fig:MOK}(c).   The same features can also be found in the Faraday rotation angles $\theta_F$ [see Fig.~\ref{fig:MOF}].  The resultant MOKE and MOFE are thus consistent with the above analysis about the electronic structures of Cr\textit{XY}. 

Especially, the peaks of $\theta_K$ and $\theta_F$ for ML and BL Cr\textit{XY} appear at almost the same incident photon energy, as shown in Figs.~\ref{fig:MOK} and~\ref{fig:MOF}.  In the moderate energy range (1 $\sim$ 5 eV), the $\theta_K$ and $\theta_F$ are remarkably large.  Taking Cr\textit{X}Br as examples, the maximal values of $\theta_K$ ($\theta_F$) for ML Cr\textit{X}Br (\textit{X} = S, Se, Te) are 0.56$^\circ$ (0.06$^\circ$), 0.54$^\circ$ (0.08$^\circ$), and 0.48$^\circ$ (0.09$^\circ$), respectively.  And the maximal values of $\theta_K$ ($\theta_F$) for BL Cr\textit{X}Br (\textit{X} = S, Se, Te) are 1.04$^\circ$ (0.11$^\circ$), 1.05$^\circ$ (0.15$^\circ$), and 0.93$^\circ$ (0.18$^\circ$), respectively.  The maximal values of $\theta_K$ and $\theta_F$ as well as the corresponding photon energies for ML and BL Cr\textit{XY} are summarized in Tab.~\ref{tab:MOKF}.  The Kerr rotation angles $\theta_K$ of ML and BL Cr\textit{XY} are comparable or even larger than that of conventional ferromagnets and some famous 2D ferromagnets, e.g., bcc Fe (-0.5$^\circ$~\cite{Guo1995}), hcp Co (-0.42$^\circ$)~\cite{Antonov2004book}, fcc Ni (-0.25$^\circ$)~\cite{Oppeneer1992}, CrI$_3$ (ML 0.29$^\circ$, trilayer 2.86$^\circ$)~\cite{Huang2017}, CrTe$_2$ (ML 0.24$^\circ$, trilayer -1.76$^\circ$)~\cite{YangXX2021}, CrGeTe$_3$ (BL 0.0007$^\circ$, trilayer 0.002$^\circ$~\cite{Gong2017}), gray arsenene (ML 0.81$^\circ$, BL-AA 0.13$^\circ$, BL-AC 0.14$^\circ$, here AA and AC are stacking patterns.)~\cite{Zhou2017}, blue phosphorene (ML 0.12$^\circ$, BL-AA 0.02$^\circ$, BL-AB 0.03$^\circ$, BL-AC 0.03$^\circ$)~\cite{Zhou2017}, InS (ML 0.34$^\circ$)~\cite{Feng2017}, and Fe$_n$GeTe$_2$ ($n$ = 3, 4, 5) (0.74$^\circ$ $\sim$ 2.07$^\circ$)~\cite{YangFGT2021}.  In addition, the Faraday rotation angles $\theta_F$ of ML and BL Cr\textit{XY} are also comparable or larger than that of some famous 2D ferromagnets, such as CrI$_3$ (ML 50$^\circ/\mu m$, BL 75$^\circ/\mu m$, trilayer 108$^\circ/\mu m$)~\cite{Kumar2019}, CrTe$_2$ (ML 30 $\sim$ -173$^\circ/\mu m$)~\cite{YangXX2021}, CrGeTe$_3$ (ML 120$^\circ/\mu m$, trilayer 100$^\circ/\mu m$)~\cite{FangY2018}, gray arsenene (ML 137.8$^\circ/\mu m$, BL-AA 9.3$^\circ/\mu m$, BL-AC 10.0$^\circ/\mu m$)~\cite{Zhou2017}, blue phosphorene (ML 18.9$^\circ/\mu m$, BL-AA 1.4$^\circ/\mu m$, BL-AB 2.3$^\circ/\mu m$, BL-AC 2.2$^\circ/\mu m$)~\cite{Zhou2017}, InS (84.3$^\circ/\mu m$)~\cite{Feng2017}, and Fe$_n$GeTe$_2$ ($n$ = 3, 4, 5) (50.28 $\sim$ 222.66$^\circ/\mu m$)~\cite{YangFGT2021}.  The above magneto-optical data are summarized in Tab.~\ref{tab:Comparison} for a better comparison.

The Kerr and Faraday rotation angles ($\theta_K$ and $\theta_F$) at a given photon energy are obviously increasing with the thin-film thickness of Cr\textit{XY}, which can be easily seen from Tab.~\ref{tab:MOKF} and Figs.~\ref{fig:MOK} and~\ref{fig:MOF}.  In order to quantitatively describe the overall trend of Kerr and Faraday spectra, we here introduce a physical quantity called magneto-optical strengths (MOS)~\cite{Feng2017,feng2020,Zhouxiaodong2021}
\begin{eqnarray}\label{eq:MOS}
\begin{split}
& \textnormal{MOS}_K = \int_{0^{+}}^\infty\hbar|\theta_K(\omega)|d\omega, \\
& \textnormal{MOS}_F = \int_{0^{+}}^\infty\hbar|\theta_F(\omega)|d\omega.
\end{split}
\end{eqnarray}
MOS$_K$ and MOS$_F$ are plotted in Figs.~\ref{fig:MOS}(a) and~\ref{fig:MOS}(b), respectively, from which one can observe that the MOKE and MOFE of BL Cr\textit{XY} are nearly double that of ML Cr\textit{XY}.  It is simply because that the MOKE and MOFE as the first-order magneto-optical effects are proportional to the magnetization, which is linearly dependent on the number of layers.  This trend has also been observed in 2D van der Waals magnets CrI$_3$~\cite{Huang2017} and Cr$_2$Ge$_2$Te$_6$~\cite{FangY2018}.  Moreover, a clear step-up trend of $\theta_K$ and $\theta_F$ can be seen in ML and BL Cr\textit{XY} by changing \textit{X} (S $\rightarrow$ Se $\rightarrow$ Te) or \textit{Y} (Cl $\rightarrow$ Br $\rightarrow$ I).  It can be understood that the MOKE and MOFE are also proportional to the SOC strength, which increases with the atomic number.  This trend can also be found in our previous work about hole-doped \textit{MX} (\textit{M} = Ga, In; \textit{X} = S, Se, Te) monolayer, in which the MOKE and MOFE have a size consequence of In$X$ $>$ Ga$X$ due to the larger SOC strength of In atom.

\begin{figure*}
	\includegraphics[width=1.7\columnwidth]{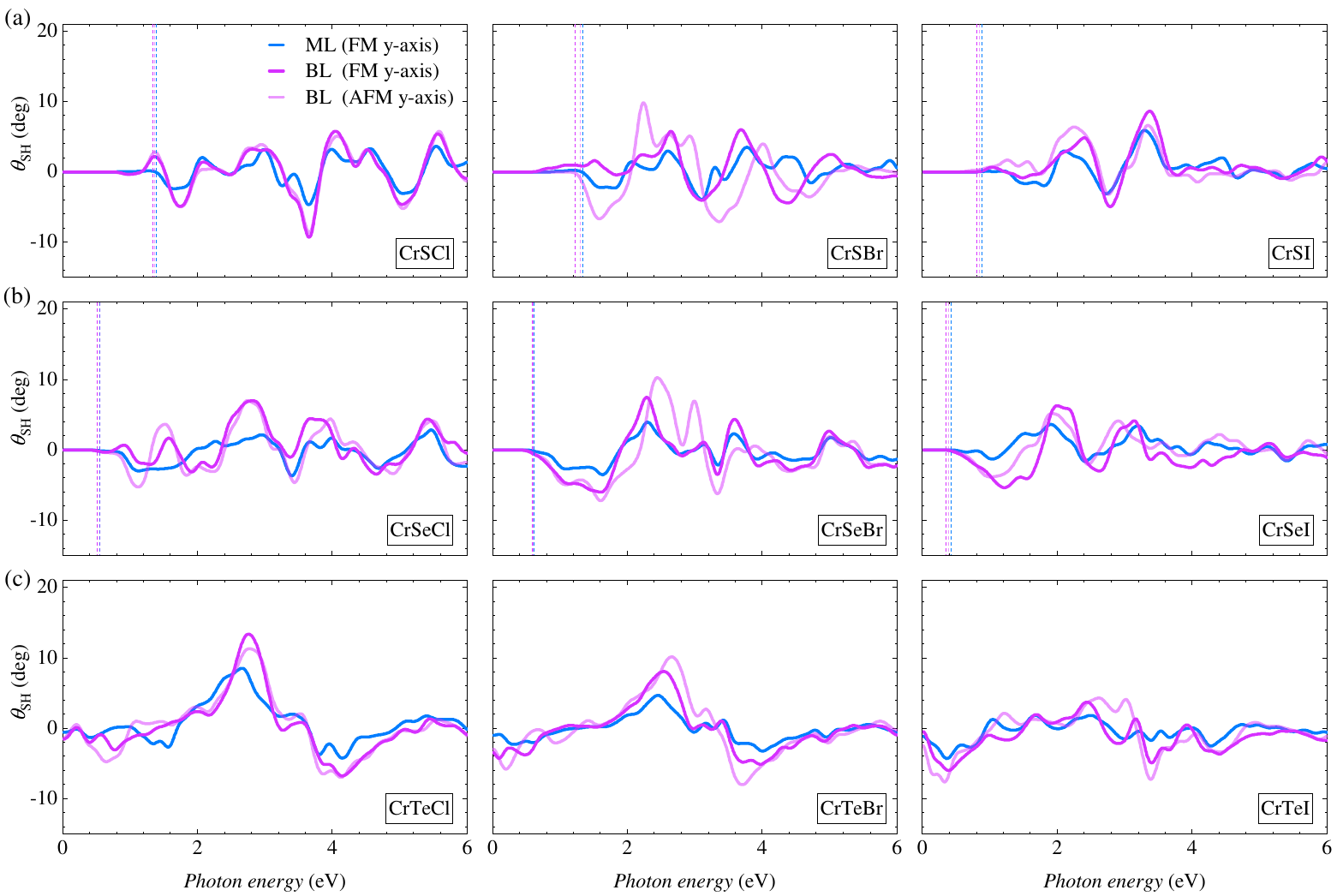}
	\caption{(Color online) The magneto-optical Sch{\"a}fer-Hubert rotation angles ($\theta_{SH}$) of monolayer ferromagnetic (blue solid lines), bilayer ferromagnetic (dark pink solid lines), and bilayer antiferromagnetic (light pink solid lines) Cr\textit{XY} with the in-plane magnetization along the $y$-axis.  The vertical dashed lines in (a) and (b) indicate the band gaps.}
	\label{fig:MOSH}
\end{figure*}

\begin{figure*}
	\includegraphics[width=1.7\columnwidth]{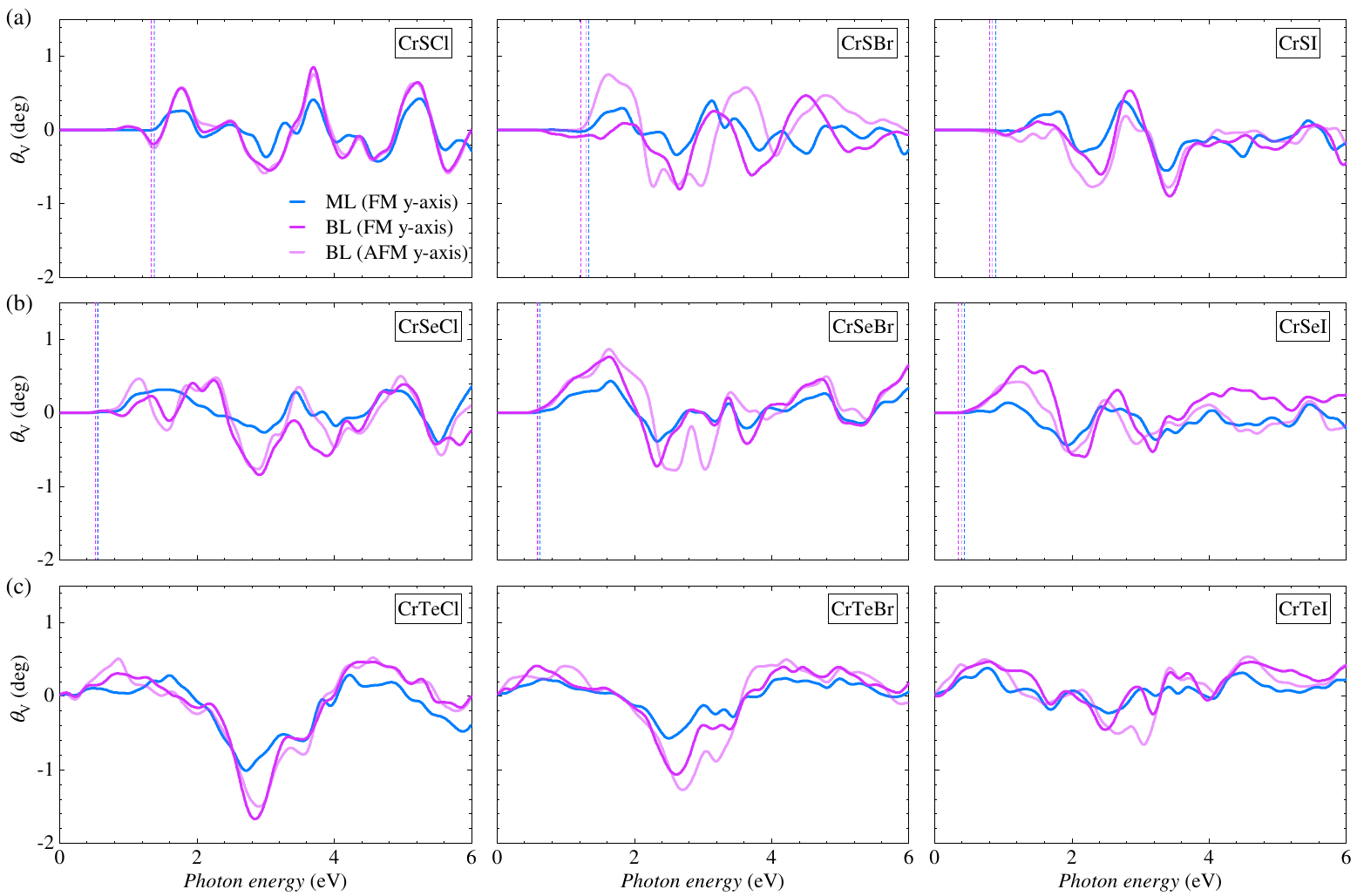}
	\caption{(Color online)  The magneto-optical Voigt rotation angles ($\theta_{V}$) of monolayer ferromagnetic (blue solid lines), bilayer ferromagnetic (dark pink solid lines), and bilayer antiferromagnetic (light pink solid lines) Cr\textit{XY} with the in-plane magnetization along the $y$-axis.  The vertical dashed lines in (a) and (b) indicate the band gaps.}
	\label{fig:MOV}
\end{figure*}

\begin{figure}
	\includegraphics[width=0.9\columnwidth]{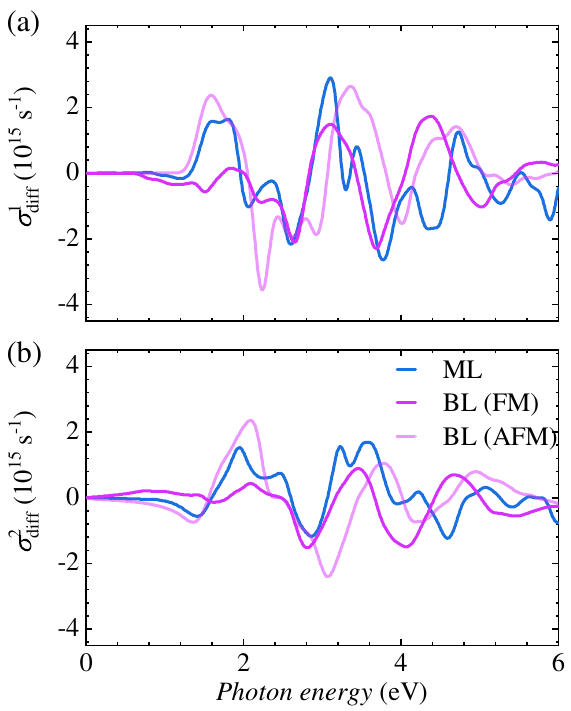}
	\caption{(Color online) The real (a) and imaginary (b) parts of the difference between the two diagonal elements of optical conductivities ($\sigma_{\textnormal{diff}}=\sigma_{yy}-\sigma_{xx}$) for monolayer ferromagnetic (blue solid lines), bilayer ferromagnetic (dark pink solid lines), and bilayer antiferromagnetic (light pink solid lines) CrSBr.}
	\label{fig:Opt}
\end{figure}

\begin{figure*}
	\includegraphics[width=2\columnwidth]{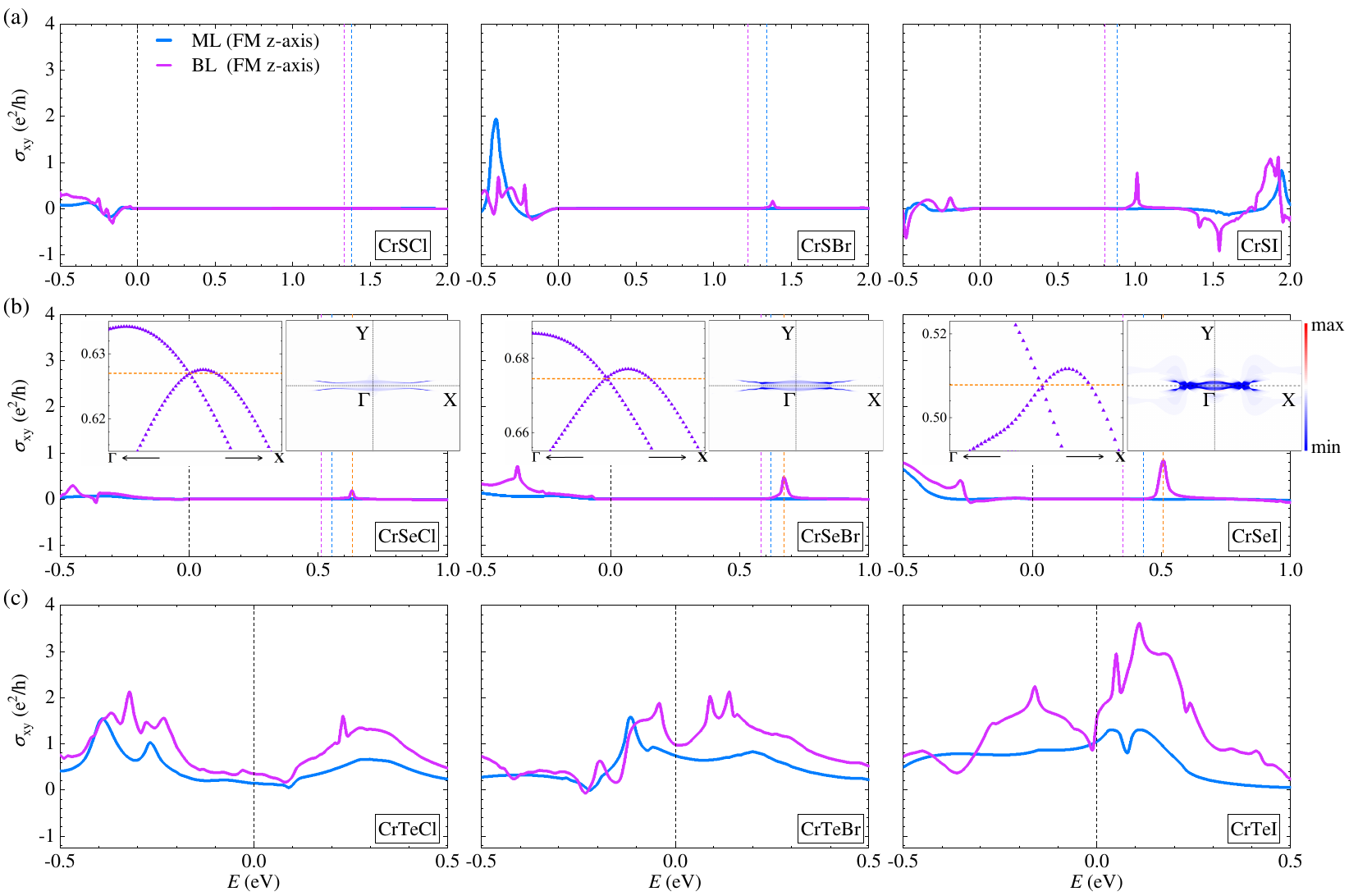}
	\caption{(Color online) The intrinsic anomalous Hall conductivities ($\sigma_{xy}$) as a function of energy for monolayer (blue solid lines) and bilayer (pink solid lines) Cr\textit{XY} with out-of-plane magnetization.  The black dashed lines indicate the true Fermi energy for metals or the top of valence bands for semiconductors.  The blue and pink dashed lines indicate the bottom of conduction bands for monolayer and bilayer semiconducting CrS\textit{Y}/CrSe\textit{Y}, respectively.  The right inset in (b) are momentum-resolved Berry curvature (in arbitrary unit) in 2D Brillouin zone.  The left inset in (b) are orbital-projected band structures, in which the violet solid triangles present the components of Cr $d_{x^2-y^2}$ orbitals and the orange dashed lines label the energy position where the Berry curvature is calculated.}
	\label{fig:AHC}
\end{figure*}

\subsection{Second-order magneto-optical effects}

The first-order magneto-optical effects (MOKE and MOFE) are forbidden in ML and BL Cr\textit{XY} with in-plane magnetization, while the second-order magneto-optical effects (MOSHE and MOVE) can arise without the restriction by symmetry.  Here, we illustrate the MOSHE and MOVE of ML FM and BL FM/AFM Cr\textit{XY} by taking the in-plane $y$-axis magnetization as an example, and the results of the $x$-axis magnetization can be obtained by putting a minus sign (refer to Eqs.~\eqref{eq:SH} and~\eqref{eq:Voigt} and note that $\epsilon_{zx}$ equals zero due to the symmetry restriction).  It has to be mentioned here that due to the in-plane crystal anisotropy of Cr\textit{XY} (a 2D rectangular primitive cell), the natural linear birefringence exists.  This phenomenon is actually not related to magnetism, and can not be mixed into the second-order magneto-optical effects~\cite{Oh1991}.  The diagonal elements of permittivity tensor can be expanded into a Taylor series in powers of magnetization \textbf{M} (up to second-order),
\begin{equation}\label{eq:permittivity}
	\epsilon_{\alpha\alpha}(\textbf{M})=\epsilon^{(0)}_{\alpha\alpha}+\epsilon^{(1)}_{\alpha\alpha} \textbf{M}+\epsilon^{(2)}_{\alpha\alpha} \textbf{M}^{2},
\end{equation}
where $\epsilon^{(0)}_{\alpha\alpha}$ is the part independent on magnetization, $\epsilon^{(1)}_{\alpha\alpha}$ and $\epsilon^{(2)}_{\alpha\alpha}$ are linearly and quadratically dependent on magnetization, respectively.  According to the Onsager relation, $\epsilon_{\alpha\alpha} (-\textbf{M})=\epsilon_{\alpha\alpha} (\textbf{M})$, we know $\epsilon^{(1)}_{\alpha\alpha}=0$.  It turns out to be clear that $\epsilon^{(0)}_{\alpha\alpha}$ is responsible for the natural linear birefringence originating from crystal anisotropy, and $\epsilon^{(2)}_{\alpha\alpha}$ is responsible for the MOSHE and MOVE originating from magnetism.  We first calculate $\epsilon^{(0)}_{\alpha\alpha}$ in 2D nonmagnetic Cr\textit{XY} by forcing the magnetic moments to be zero during self-consistent field calculation, and then calculate $\epsilon_{\alpha\alpha}$ (the sum of $\epsilon^{(0)}_{\alpha\alpha}$ and $\epsilon^{(2)}_{\alpha\alpha}$) in 2D magnetic Cr\textit{XY}.   Finally, $\epsilon^{(2)}_{\alpha\alpha}$ can be simply obtained by subtracting $\epsilon^{(0)}_{\alpha\alpha}$ from $\epsilon_{\alpha\alpha}$.  Thus, we take $\epsilon^{(2)}_{\alpha\alpha}$ instead of $\epsilon_{\alpha\alpha}$ into Eqs.~\eqref{eq:SH} and~\eqref{eq:Voigt}.  The corresponding optical conductivities $\sigma^{(2)}_{\alpha\alpha}$ are shown in supplemental Figs.~\textcolor{blue}{S16} and~\textcolor{blue}{S17}~\cite{SuppMater}.

The calculated Sch{\"a}fer-Hubert and Voigt rotation angles ($\theta_{SH}$ and $\theta_{V}$) of Cr\textit{XY} are plotted in Figs.~\ref{fig:MOSH} and~\ref{fig:MOV}, and the corresponding ellipticities ($\varepsilon_{SH}$ and $\varepsilon_{V}$) are plotted in supplemental Figs.~\textcolor{blue}{S18} and~\textcolor{blue}{S19}~\cite{SuppMater}.  The spectra of $\theta_{SH}$ and $\theta_{V}$ for semiconducting CrS\textit{Y} and CrSe\textit{Y} are negligibly small in the low energy range far below the band gaps.  However, in contrast to the Kerr and Faraday effects, $\theta_{SH}$ and $\theta_{V}$ take on finite values near the band edges [e.g., Figs.~\ref{fig:MOSH}(a) and \ref{fig:MOV}(a)].  The reason is that $\theta_{SH}$ and $\theta_{V}$ depend on the diagonal elements of optical conductivity, which are nonvanishing within the band gap (see Figs.~\textcolor{blue}{S16} and~\textcolor{blue}{S17}~\cite{SuppMater}).  In the energy range larger than the band gaps, $\theta_{SH}$ and $\theta_{V}$ oscillate frequently, as shown in Figs.~\ref{fig:MOSH}(a,b) and \ref{fig:MOV}(a,b), respectively.  For metallic CrTe\textit{Y}, the relatively larger $\theta_{SH}$ and $\theta_{V}$ can be found in Figs.~\ref{fig:MOSH}(c) and \ref{fig:MOV}(c), respectively.   Taking CrSBr as an example, the maximal values of $\theta_{SH}$ for ML FM, BL FM, and BL AFM states are -3.92$^\circ$, 6.01$^\circ$, and 9.87$^\circ$ at the photon energies of 3.10 eV, 3.69 eV, and 2.24 eV, respectively; the maximal values of $\theta_{V}$ for ML FM, BL FM, and BL AFM states are 0.40$^\circ$, -0.80$^\circ$, and -0.77$^\circ$ at the photon energies of 3.13 eV, 2.66 eV, and 2.28 eV, respectively.  The data of the largest $\theta_{SH}$ and $\theta_{V}$ as well as the related photon energies for all Cr\textit{XY} family members are summarized in Tab.~\ref{tab:MOKF}.  The $\theta_{SH}$ and $\theta_{V}$ of Cr\textit{XY} are larger than that of some 2D AFMs, for example, 10 nm thick CoO thin-film ($\theta_{SH} \sim$ 0.007 deg)~\cite{ZhengZ2018} and 10 nm thick CuMnAs thin-film ($\theta_{V} \sim$ 0.023 deg)~\cite{Saidl2017}.

From Eqs.~\eqref{eq:SH} and~\eqref{eq:Voigt}, one can know that $\theta_{SH}$, $\varepsilon_{SH}$, $\theta_{V}$, and $\varepsilon_{V}$ are dominated by the two diagonal elements of permittivity tensor ($\epsilon_{yy}$ and $\epsilon_{xx}$) as the off-diagonal element ($\epsilon_{zx}$) is restricted to be zero by symmetry.  By plugging the optical conductivity ($\epsilon_{yy(xx)}=1+\frac{4\pi i}{\omega}\sigma_{yy(xx)}$) into Eq.~\eqref{eq:SH}, one shall find $\theta_{SH}\propto -\sigma^{1}_{\textnormal{diff}}=-\textnormal{Re}(\sigma_{yy}-\sigma_{xx}$) and $\varepsilon_{SH}\propto -\sigma^{2}_{\textnormal{diff}}=-\textnormal{Im}(\sigma_{yy}-\sigma_{xx}$).  Figure~\ref{fig:Opt} plots the real ($\sigma_\textnormal{diff}^1$) and imaginary ($\sigma_\textnormal{diff}^2$) parts of the difference between the diagonal elements of optical conductivities ($\sigma_{yy}$ and $\sigma_{xx}$) for ML FM, BL FM, and BL AFM CrSBr.  One can clearly observe that the spectral structure of $\theta_{SH}$ is determined by $\sigma_\textnormal{diff}^1$ only differing by an opposite sign, comparing the middle panel of Fig.~\ref{fig:MOSH}(a) with Fig.~\ref{fig:Opt}(a).  The spectral structure of $\varepsilon_{SH}$ is thus determined by $\sigma_\textnormal{diff}^2$, comparing the middle panel of supplemental Fig.~\textcolor{blue}{S18}(a)~\cite{SuppMater} with Fig.~\ref{fig:Opt}(b).  On the other hand, there is not such a clear relation between $\theta_{V}$ ($\varepsilon_{V}$) and $\sigma_\textnormal{diff}^1$ ($\sigma_\textnormal{diff}^2$) since equation~\eqref{eq:Voigt} involves the square roots of $\epsilon_{yy}$ and $\epsilon_{xx}$, though the spectral structures of $\theta_{V}$ ($\varepsilon_{V}$) and $\sigma_\textnormal{diff}^1$ ($\sigma_\textnormal{diff}^2$) seem to be similar somewhere.

\subsection{Anomalous Hall, anomalous Nernst, and anomalous thermal Hall effects}
In this subsection, we shall discuss the AHE as well as its thermoelectric counterpart, ANE, and its thermal analogue, ATHE, for ML and BL FM Cr\textit{XY} with out-of-plane magnetization.  The ANE and ATHE are evaluated at 100 K below the Curie temperatures by the theoretical prediction ($>$ 108 K)~\cite{WANG2019293} and experimental measurements ($>$ 140 K)~\cite{Nathan2021,LeeKH2021}.  The magnetic transition temperature can be in principle increased via doping effects.  For example, magnetic atom-doped 2D transition metal dichalcogenides have been predicted to exhibit room-temperature ferromagnetism~\cite{Ashwin2013,SunLL2016}.  Additionally, the Curie temperature of trilayer Fe$_3$GeTe$_2$ can be improved to room-temperature by ionic-liquid gating technique~\cite{Deng2018}.

The calculated intrinsic AHC ($\sigma_{xy}$) as a function of energy are displayed in Fig.~\ref{fig:AHC}.  For semiconducting CrS\textit{Y} and CrSe\textit{Y}, the AHC is zero within the band gaps, similarly to the magneto-optical Kerr and Faraday effects.  For metallic CrTe\textit{Y} (\textit{Y} = Cl, Br, I), $\sigma_{xy}$ takes the values of 0.14 $e^2/h$, 0.75 $e^2/h$, and 1.05 $e^2/h$ for ML FM states, and of 0.35 $e^2/h$, 0.98 $e^2/h$, and 1.56 $e^2/h$ for BL FM states at the Fermi energy ($E_F$), respectively.  The magnetotransport properties like the AHE can be modulated by electron or hole doping, which shifts the $E_F$ upward or downward, respectively.  According to the recent experimental works~\cite{LiLJ2016,Deng2018}, the doping carrier concentration for 2D materials can reach up to 10$^{15}$ cm$^{-2}$ via the gate voltage.  Then, the variation range of the $E_F$ can be estimated in a reasonable range, see supplemental Fig.~\textcolor{blue}{S20}~\cite{SuppMater}, by taking ML Cr\textit{X}Br as examples.  The energy ranges for ML and BL CrS\textit{Y}, CrSe\textit{Y}, and CrTe\textit{Y} are reasonably given to be [-0.5, 2.0] eV, [-0.5, 1.0], and [-0.5, 0.5] eV, respectively.  In the above energy regions, $\sigma_{xy}$ increases up to 1.93 $e^2/h$, 1.54 $e^2/h$, and 1.57 $e^2/h$ for ML CrSBr, CrTeCl, and CrTeBr at -0.40 eV, -0.39 eV, and -0.12 eV by hole doping, respectively.  And $\sigma_{xy}$ increases up to 2.12 $e^2/h$, 1.88 $e^2/h$, and 2.24 $e^2/h$ for BL CrTeCl, CrTeBr, and CrTeI  at -0.32 eV, -0.04 eV, and -0.16 eV by hole doping, respectively.  On the other hand for electron doping, $\sigma_{xy}$ can be reached to 1.60 $e^2/h$, 2.12 $e^2/h$, 3.61 $e^2/h$ for BL CrTeCl, CrTeBr, and CrTeI at 0.23 eV, 0.14 eV, and 0.11 eV, respectively.  It is obvious that the metallic CrTe\textit{Y} exhibit more prominent AHE than the semiconducting CrS\textit{Y} and CrSe\textit{Y}.  The largest $\sigma_{xy}$ for CrTe\textit{Y} are in the range of 544.46--827.95 S/cm (converted from the unit of $e^2/h$ by considering the effective thickness), which is larger than bulk Fe$_3$GeTe$_2$ (140--540 S/cm)~\cite{KimKyoo2018}, thin film Fe$_3$GeTe$_2$ ($\sim$ 400 S/cm)~\cite{Xu2019}, thin film Fe$_4$GeTe$_2$ ($\sim$ 180 S/cm)~\cite{Seo2020}, and is even compared with bulk bcc Fe (751 S/cm~\cite{Yao2004}, 1032 S/cm~\cite{Dheer1967}).

Strikingly, a peak of $\sigma_{xy}$ appears in BL CrS\textit{Y} and CrSe\textit{Y} near the bottom of conduction bands, and grows larger with the increasing of atomic number (Cl $\rightarrow$ Br $\rightarrow$ I), as displayed in Figs.~\ref{fig:AHC}(a) and~\ref{fig:AHC}(b).  We interpret the underlying physical reasons by taking BL CrSe\textit{Y} as examples.  At the energy position of the peaked $\sigma_{xy}$, there exists a band crossing in the band structure that is dominated by Cr $d_{x^2-y^2}$ orbitals [see the left inset in Fig.~\ref{fig:AHC}(b)].  The band crossing is a Weyl-like point, which is protected by the glide symmetry $\mathcal{M}_z\tau_{(\frac{1}{2},\frac{1}{2},0)}$.  It preserves even after including SOC because the glide symmetry still exists.  Pinning the $E_F$ at the band crossing point, the Berry curvature exhibits hot spots around the $\Gamma$ point [see the right inset in Fig.~\ref{fig:AHC}(b)], which definitely gives rise to a large $\sigma_{xy}$.  With the increasing of atomic number $Y$, the region of the hot spots of Berry curvature enlarges gradually, which enhances the peak of $\sigma_{xy}$.

\begin{figure*}
	\includegraphics[width=1.6\columnwidth]{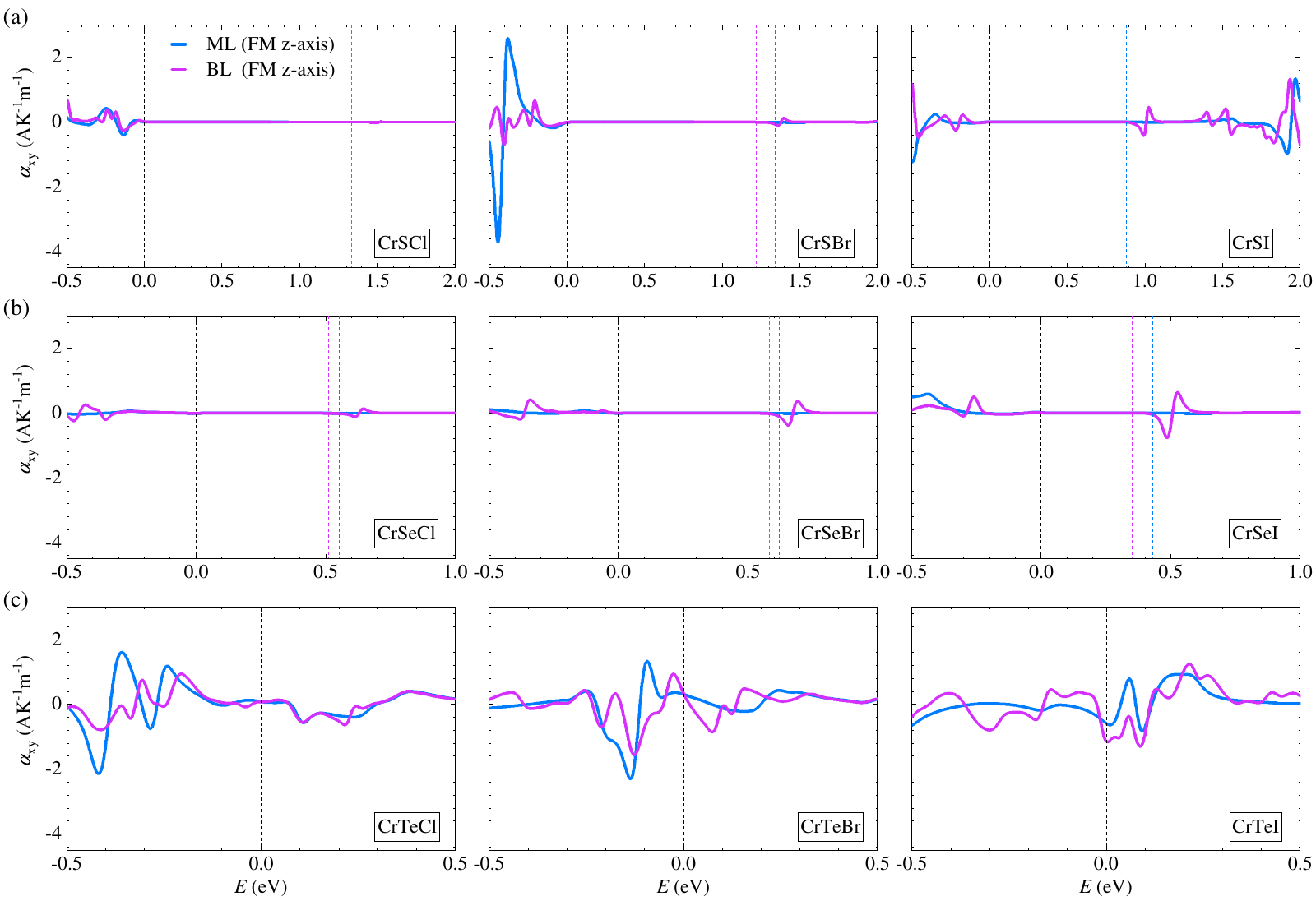}
	\caption{(Color online) The anomalous Nernst conductivities ($\alpha_{xy}$) calculated at 100 K as a function of energy for monolayer (blue solid lines) and bilayer (pink solid lines) Cr\textit{XY} with out-of-plane magnetization.  The black dashed lines indicate the true Fermi energy for metals or the top of valence bands for semiconductors.  The blue and pink dashed lines indicate the bottom of conduction bands for monolayer and bilayer semiconducting CrS\textit{Y}/CrSe\textit{Y}, respectively.}
	\label{fig:ANC}
\end{figure*}

\begin{figure*}
	\includegraphics[width=1.6\columnwidth]{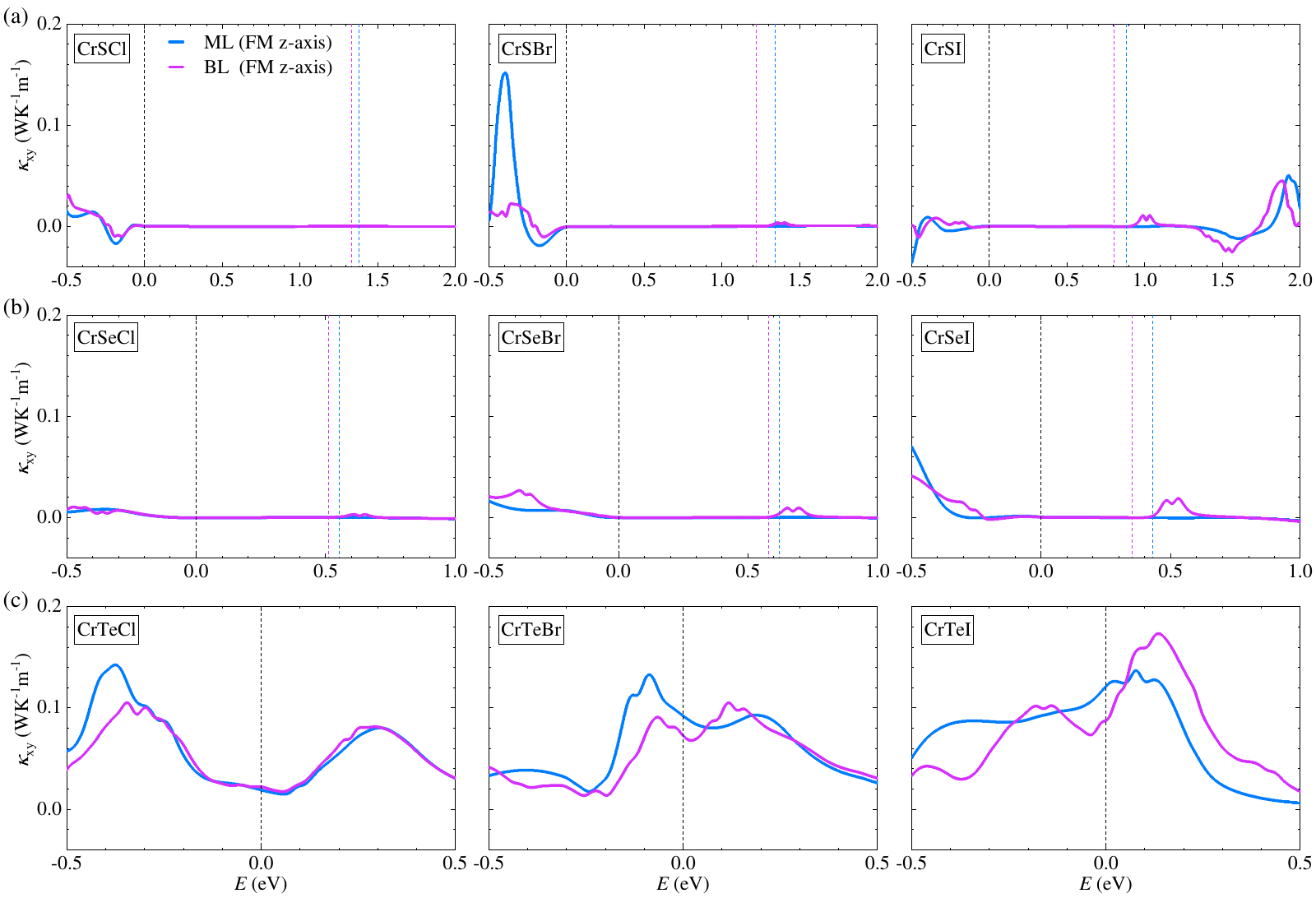}
	\caption{(Color online) The anomalous thermal Hall conductivities ($\kappa_{xy}$) calculated at 100 K as a function of energy for monolayer (blue solid lines) and bilayer (pink solid lines) Cr\textit{XY} with out-of-plane magnetization.  The black dashed lines indicate the true Fermi energy for metals or the top of valence bands for semiconductors.  The blue and pink dashed lines indicate the bottom of conduction bands for monolayer and bilayer semiconducting CrS\textit{Y}/CrSe\textit{Y}, respectively.}
	\label{fig:ATHC}
\end{figure*}

The ANE, the thermoelectric counterpart of AHE,  is a celebrated effect from the realm of the spin caloritronics~\cite{Bauer2012,Boona2014}.  At the low temperature, the ANC ($\alpha_{xy}$) is connected to the energy derivative of $\sigma_{xy}$ via the Mott relation~\cite{Xiao2006},
\begin{eqnarray}\label{eq:MottANC}
\alpha_{xy}&=&-\frac{\pi^2k_B^2T}{3e}\sigma^{\prime}_{xy}(E),
\end{eqnarray}
which can be derived from Eq.~\eqref{eq:ANC}.  From the above equation, one can find that a steep slope of $\sigma_{xy}$ versus energy generally leads to a large $\alpha_{xy}$.  For example, there appear two opposite peaks of $\alpha_{xy}$ close to the bottom of conduction bands for BL CrS\textit{Y} and CrSe\textit{Y} [see Figs.~\ref{fig:ANC}(a) and~\ref{fig:ANC}(b)], which exactly originates from the positive and negative slopes of the peak of $\sigma_{xy}$ at the same energy [see Fig.~\ref{fig:AHC}(a) and(b)].  For semiconducting CrS\textit{Y} and CrSe\textit{Y}, $\alpha_{xy}$ is unsurprisingly zero within the electronic band gap.  Among CrS\textit{Y} and CrSe\textit{Y}, ML CrSBr exhibits the excellent ANE under appropriate hole doping, such as 2.59 AK$^{-1}$m$^{-1}$ and -3.71 AK$^{-1}$m$^{-1}$ at -0.38 eV and -0.44 eV, respectively.  For metallic ML (BL) CrTe\textit{Y} with $Y=$ Cl, Br, and I, the calculated $\alpha_{xy}$ at the true Fermi energy are 0.09 (0.09) AK$^{-1}$m$^{-1}$, 0.32 (0.36) AK$^{-1}$m$^{-1}$, and -0.58 (-1.13) AK$^{-1}$m$^{-1}$, respectively.  By doping holes or electrons, $\alpha_{xy}$ can reach up to -2.13 (0.94) AK$^{-1}$m$^{-1}$ at -0.42 (-0.20) eV for ML (BL) CrTeCl, to -2.29 (-1.55) AK$^{-1}$m$^{-1}$ at -0.14 (-0.12) eV for ML (BL) CrTeBr, and to 0.90 (-1.29) AK$^{-1}$m$^{-1}$ at 0.16 (0.09) eV for ML (BL) CrTeI, respectively.  The calculated $|\alpha_{xy}|$ of ML CrSBr, CrTeCl, and CrTeBr are considerably large ($>$ 2.0 AK$^{-1}$m$^{-1}$), which exceeds to bulk ferromagnets Co$_3$Sn$_2$S$_2$ ($\sim$ 2.0 AK$^{-1}$m$^{-1}$)~\cite{Guin2019} and Ti$_2$MnAl ($\sim$ 1.31 AK$^{-1}$m$^{-1}$)~\cite{Noky2018}, and even comparable with multi-layer Fe$_n$GeTe$_2$ ($n$ = 3, 4, 5) (0.3--3.3 AK$^{-1}$m$^{-1}$)~\cite{YangFGT2021}, suggesting that 2D Cr\textit{XY} is an excellent material platform for thermoelectric applications.

The ATHE, the thermal analogue of AHE, has been widely used to study the charge neutral quasiparticles in quantum materials~\cite{Onose2010,Max2015,Max2015_2,Doki2018,Sugii2017,Akazawa2020,Zhang2021}.
The calculated ATHC ($\kappa_{xy}$) of ML and BL Cr\textit{XY} are plotted in Fig.~\ref{fig:ATHC}.  Similarly to $\sigma_{xy}$ and $\alpha_{xy}$, $\kappa_{xy}$ is vanishing within the band gaps of CrS\textit{Y} and CrSe\textit{Y}.  The calculated $\kappa_{xy}$ at the true Fermi energy for ML (BL) CrTe\textit{Y} with $Y=$ Cl, Br, and I are 0.02 (0.02) WK$^{-1}$m$^{-1}$, 0.09 (0.07) WK$^{-1}$m$^{-1}$, and 0.12 (0.09) WK$^{-1}$m$^{-1}$, respectively.  The ATHE can be further enhanced by doping appropriate holes or electrons.  For instance, $\kappa_{xy}$ increases to its largest value of 0.15 WK$^{-1}$m$^{-1}$ at -0.39 eV for ML CrSBr.  In the case of ML (BL) CrTe\textit{Y}, the largest $\kappa_{xy}$ reaches up to 0.14 (0.11) WK$^{-1}$m$^{-1}$, 0.13 (0.10) WK$^{-1}$m$^{-1}$, and 0.14 (0.17) WK$^{-1}$m$^{-1}$ at -0.37 (-0.34) eV, -0.09 (0.11) eV, and 0.08 (0.13) eV, respectively.  The $\kappa_{xy}$ of 2D Cr\textit{XY} are much larger than that of other vdW ferromagnets, e.g., VI$_3$ (0.01 WK$^{-1}$m$^{-1}$)~\cite{Zhang2021}.

\section{Summary}

In summary, we systematically investigated the band structures, magnetocrystalline anisotropy energy, first- and second-order magneto-optical effects, and intrinsically anomalous transport properties in monolayer and bilayer Cr\textit{XY} (\textit{X} = S, Se, Te; \textit{Y} = Cl, Br, I) by employing the first-principles calculations.  From the band structures, monolayer and bilayer CrS\textit{Y} and CrSe\textit{Y} are identified to be narrow band-gap semiconductors, whereas monolayer and bilayer CrTe\textit{Y} are multi-band metals.  The results of magnetocrystalline anisotropy energy show that CrTeBr and CrTeI prefer to out-of-plane magnetization, whereas the other seven family members of Cr\textit{XY} are in favor of in-plane magnetization.  The magnetic ground states for bilayer Cr\textit{XY} are confirmed, that is, bilayer CrSI, CrSeCl, and CrTeCl present strong interlayer ferromagnetic coupling, whereas the other ones show interlayer antiferromagnetic coupling. The magnetic group theory identifies the nonzero elements of optical conductivity tensor, which in turn determines the symmetry of magneto-optical Kerr and Faraday effects as well as the anomalous Hall, anomalous Nernst, and anomalous thermal Hall effects.  The above physical effects can only exist in monolayer and bilayer ferromagnetic Cr\textit{XY} with out-of-plane magnetization.  The magneto-optical strength for the Kerr and Faraday effects of bilayer Cr\textit{XY} are nearly two times that of monolayer Cr\textit{XY} due to the doubled magnetization, and the magneto-optical strength of both monolayer and bilayer Cr\textit{XY} increase distinctly with the increasing of atomic number in \textit{X} or \textit{Y} because of the enhanced spin-orbital coupling.  The largest Kerr and Faraday rotation angles in the moderate energy range (1 $\sim$ 5 eV) are found in bilayer CrSeI, which turns out to be 1.49$^\circ$ and 0.20$^\circ$, respectively.  The anomalous transport properties of Cr\textit{XY} are considerably large under proper hole or electron doping.  For instance, the anomalous Hall, anomalous Nernst, and anomalous thermal Hall conductivities reach up to their maximums of 3.61 $e^2/h$, -3.71 AK$^{-1}$m$^{-1}$, and 0.17 WK$^{-1}$m$^{-1}$ in bilayer CrTeI, monolayer CrSBr, and bilayer CrTeI, respectively.  In the case of in-plane magnetization, the second-order magneto-optical effects do exist in both ferromagnetic and antiferromagnetic Cr\textit{XY}, in contrast to the first-order ones.  The calculated Sch{\"a}fer-Hubert and Voigt effects are considerable large in Cr\textit{XY}.  For example, the largest Sch{\"a}fer-Hubert and Voigt rotation angles are found to be 13.39 $^\circ$ and -1.67$^\circ$ in bilayer ferromagnetic CrTeCl, and to be 11.31$^\circ$ and -1.50$^\circ$ in bilayer antiferromagnetic CrTeCl, respectively.

Our results show that 2D van der Waals layered magnets Cr\textit{XY} may be excellent material candidates for promising applications to magneto-optical devices, spintronics, and spin caloritronics.  For example, the semiconducting Cr\textit{XY} can be used to make van der Waals magnetic heterostructures, similarly to semiconducting ferromagnetic CrI$_3$, which realizes the quantum anomalous Hall effect~\cite{ZhangJY2018} and valley manipulation~\cite{ZhongD2017,Seyler2018}.  The metallic Cr\textit{XY} can also be used to make heterostructures, similarly to metallic Fe$_3$GeTe$_2$, which can increase the critical temperatures of 2D magnetic materials~\cite{WangHY2020}, fabricate giant magnetoresistance (GMR) devices~\cite{Albarakati2019}, design metal electrodes~\cite{WuQy2019}, and tailor the anomalous transport properties~\cite{ShaoYan2020}.

\begin{acknowledgments}
The authors thank Gui-Bin Liu, Chaoxi Cui, and Shifeng Qian for their helpful discussion.  This work is supported by the National Natural Science Foundation of China (Grant Nos. 11874085, 11734003, and 12061131002), the Sino-German Mobility Programme (Grant No. M-0142), the National Key R\&D Program of China (Grant No. 2020YFA0308800), and the Science \& Technology Innovation Program of Beijing Institute of Technology (Grant No. 2021CX01020).
\end{acknowledgments}
	
\bibliography{references}

\end{document}